\documentclass[twocolumn]{aastex631}

\usepackage{amsmath}	

\usepackage{fontspec}
\usepackage{unicode-math}
\usepackage{verbatim}
[   Extension={.otf},
    Path=./STIX2fonts/,
    Scale=1]
[   range=\mathit/{latin,Latin,greek,Greek},
    Extension = {.otf},
    BoldFont = XITSMath-Bold]

\usepackage{hyperref}
\usepackage{placeins}

\usepackage{calc}

\newcommand{\be}{\begin{equation}}
\newcommand{\ee}{\end{equation}}
\newcommand{\beal}{\begin{aligned}}
\newcommand{\eeal}{\end{aligned}}

\newcommand{\lowmdot}{\texttt{sim-d36}}
\newcommand{\highmdot}{\texttt{sim-d100}}
\newcommand{\Lx}{\ensuremath{L_\mathrm{X}}}
\newcommand{\Rm}{\ensuremath{R_\mathrm{m}}}
\newcommand{\Rlc}{\ensuremath{R_\mathrm{LC}}}

\usepackage{xcolor}

\begin{document}

\title{Modes in Transitional Millisecond Pulsars: \\ Evidence of Pulsar Wind-Induced Disk Heating from GRMHD and Radiative Transfer}

\author[0000-0002-3072-1496]{Raphaël Mignon-Risse}
\affiliation{Department of Physics, 
Norwegian University of Science and Technology,
NO-7491 Trondheim, Norway}
\affiliation{Department of Theoretical Physics, Atomic and Optics, University of Valladolid UVA, 47011 Valladolid, Spain}
\affiliation{Aix Marseille Univ, CNRS, CNES, LAM, Marseille, France}

\author[0000-0002-0237-1636]{Manuel Linares}
\affiliation{Department of Physics, Norwegian University of Science and Technology, NO-7491 Trondheim, Norway}
\affiliation{Departament de Física, EEBE, Universitat Politècnica de Catalunya, Av. Eduard Maristany 16, E-08019 Barcelona, Spain}

\author[0000-0001-6173-0099]{Kyle Parfrey}
\affiliation{Princeton Plasma Physics Laboratory, Princeton, NJ 08540, USA}

\author[0000-0002-9182-2047]{Alexander Tchekhovskoy}
\affiliation{Center for Interdisciplinary Exploration \& Research in Astrophysics (CIERA), Physics \& Astronomy, Northwestern University, Evanston, IL 60201, USA}
\affiliation{NSF-Simons AI Institute for the Sky (SkAI), 172 E. Chestnut St., Chicago, Illinois 60611, USA}

\author[0000-0003-0220-5723]{Sean Ressler}
\affiliation{Canadian Institute for Theoretical Astrophysics, University of Toronto, Toronto, ON, Canada M5S 3H8}



\begin{abstract}
Transitional millisecond pulsars (tMSPs) alternate between radio and X-ray pulsar states, 
and can represent the missing link between 
rotation- and accretion-powered neutron stars.
Their \emph{disk state} switches 
stochastically between the low and high X-ray modes, both of unknown physical origin and less luminous than low-mass X-ray binaries. 
To reveal the source of the X-ray emission, we carry out 2D axisymmetric general-relativistic magnetohydrodynamical simulations of the interaction between an accretion disk and tMSP magnetosphere.
For the first time, we post-process tMSP simulations with a radiative transfer code that incorporates thermal synchrotron, absorption, and Compton scattering processes.
By varying the disk density, hence the inflow rate, we explore two disk regimes: one truncated outside and another inside the light cylinder.
In the former, most of the X-ray flux comes from the synchrotron emission powered by the wind heating the disk: this {\lq}wind{\rq} regime could correspond to the \emph{high X-ray mode}. 
The latter is the propeller regime and lacks this heating process.
However, the propeller episodically expels the disk, activating the wind heating: a $70\%$--$30\%$ mixture of such propeller and wind regimes reproduces the X-ray spectrum of the \emph{low X-ray mode}.
%
The excess electromagnetic torque in the propeller regime increases the spin-down rate, averaged over both modes, by a few percent above the disk-free radio pulsar state, in agreement with observations.
%
Overall, the system is more luminous in X-rays when the flow is truncated {\it outside} the light cylinder and supports a contribution from wind--induced disk heating in both low and high X-ray modes.
\end{abstract}

\section{Introduction} 
\label{sec:intro}

Transitional millisecond pulsars (tMSPs) are a rare subclass of neutron stars in binary systems, with three confirmed and about eight candidate systems (in the Galactic field) known to date (\citealt{papitto_transitional_2022,koljonen_spidercat_2025}\footnote{See the online catalog at \href{https://astro.phys.ntnu.no/SpiderCAT/}{https://astro.phys.ntnu.no/SpiderCAT/}.}).
They can be observed in the radio pulsar state and in the disk or low-mass X-ray binary (LMXB) state.
They are often considered to be the evolutionary bridge between accreting and radio millisecond pulsars \citep{alpar_new_1982,radhakrishnan_origin_1982} and an observational confirmation of the ``recycling scenario,'' which explains the neutron star spin-up to millisecond periods via Gyr-long phases of accretion within LMXBs  \citep{alpar_new_1982}.
The three confirmed tMSPs to date are: PSR J1023+0038 (\citealt{archibald_radio_2009,stappers_state_2014}, J1023 hereafter), IGR J18245-2452 \citep{papitto_swings_2013} and XSS J12270-4859 \citep{bassa_state_2014}.

 In contrast to LMXBs (with X-ray luminosity up to $\Lx\, {\sim} \,10^{38} \, \mathrm{erg\, s^{-1}}$; 0.5--10~keV),
tMSPs 
exhibit a surprisingly faint accretion disk state ($\Lx\, {\sim} \,10^{33-34} \, \mathrm{erg\, s^{-1}}$) that is hence also dubbed the \textit{sub-luminous} disk state.
The disk state can be further divided into the \textit{high} (or \textit{active}) and \textit{low} (or \textit{passive}) X-ray modes, with sporadic flares \citep{linares_x-ray_2014,bogdanov_coordinated_2015}. 
We will refer to these as low- or low-$\Lx$ modes, and high- or high-$\Lx$ modes, respectively, without distinction. 
The transitions between the low and high modes are associated with variations by a factor of ${\sim}\,5{-}10$ in 
$\Lx$ \citep{linares_neutron_2014}.
It is unclear whether the multi-wavelength emission in this state is accretion-powered, rotation-powered, or both,
because both sources of energy are plausible.
Indeed, the rotational energy extraction rate, or spin-down luminosity, is estimated to be ${\sim} 10^{33-35} \, \mathrm{erg \, s^{-1}}$ among compact binary MSPs (e.g., \citealt{archibald_long-term_2013,roy_discovery_2015,koljonen_spidercat_2025}).
Furthermore, the sudden (less than tens of seconds, \citealt{baglio_matter_2023}) mode-switching behavior is not understood.

Despite the evidence of an accretion disk (broad, double-peaked optical emission lines, see, e.g., \citealt{bond_first_2002}, \citealt{linares_psr_2014}, \citealt{shahbaz_binary_2019}), the non-thermal nature of the X-ray spectrum (e.g., \citealt{linares_x-ray_2014}) rules out a standard thermal disk model \citep{shakura_black_1973} for the X-ray emission.
Additionally, X-ray pulsations at the pulsar spin period are detected only in the high mode (\citealt{archibald_accretion-powered_2015}, \citealt{papitto_x-ray_2015}).
Mode switching, although initially discovered in X-rays, correlates with flux variations across the electromagnetic spectrum: X-ray and Ultraviolet (UV) lightcurves positively correlate (e.g.,
\citealt{baglio_matter_2023}), whereas X-ray and radio fluxes anti-correlate (\citealt{bogdanov_simultaneous_2018}, \citealt{baglio_matter_2023}). 

\looseness=-1
A natural parameter to explain this rich variability,
for a given neutron star (NS) magnetic field strength,
is the (variable) mass accretion/inflow rate.
In analytical models,
this sets the position of the disk's inner edge with respect to the characteristic radii of the system \citep{1973ApJ...179..585D,1977ApJ...217..578G}.
In tMSPs the inner disk is thought to be close to the
light cylinder (e.g., \citealt{linares_neutron_2014,parfrey_general-relativistic_2017,linares_x-ray_2022}), which is defined as the surface 
at which 
a hypothetical object corotating with the star 
would reach the speed of light.
Switches between magnetically channeled accretion 
in the high mode and wind--disk shock emission in the low mode have been proposed 
to explain this rapid moding behavior 
(\citealt{linares_neutron_2014,campana_physical_2016,coti_zelati_simultaneous_2018}).
Alternatively, the disk could be located just outside the light cylinder in the high-$\Lx$ mode, enhancing the pulsar wind--disk collision and triggering X-ray synchrotron emission (as proposed by \citealt{papitto_pulsating_2019} and \citealt{veledina_pulsar_2019}).
In this scenario, the low-$\Lx$ mode could result either from the disk penetrating the light cylinder, i.e., the propeller regime \citep{veledina_pulsar_2019}, or being located even further away from the pulsar \citep{papitto_pulsating_2019,campana_probing_2019}.

For J1023, \cite{jaodand_timing_2016} and \cite{burtovoi_spin-down_2020} reported disk state spin-down rates (i.e., averaged over both low and high modes) that are only ${\sim}27\%$ and ${\sim}5\%$ higher than the radio pulsar state, respectively,  
disfavoring accretion onto the NS surface.
Similarly, the theoretical accretion rate estimated from the X-ray luminosity---assuming, for simplicity, full conversion of accretion energy---is so low that it is consistent with a flow
unable to accrete inward against the pulsar wind.
\cite{ambrosino_optical_2017} detected optical pulsations, which
time-delay studies \citep{illiano_investigating_2023} and polarization measurements \citep{baglio_polarized_2025} suggest originate from the same region as the X-rays, outside the light cylinder.
UV pulsations were reported as well (e.g., \citealt{miraval_zanon_uv_2022}), and the pulsed flux distribution is consistent with a single power law extending from the optical to X-ray bands \citep{miraval_zanon_uv_2022,baglio_polarized_2025}, indicating a common origin.

In this Letter, we investigate 
the origin, multi-wavelength emission, and torques of the low and high modes in tMSPs.
For this, we model the interaction between the pulsar magnetosphere and accretion flow 
in general-relativistic (GR) magnetohydrodynamics (MHD; see, e.g., \citealt{miller_magnetohydrodynamic_1997}, \citealt{parfrey_general-relativistic_2017}, \citealt{das_grmhd_2022}) 
for a wide range of gas inflow rates.
%
We begin with a presentation of our numerical approach in Sec.~\ref{sec:methods} and of our particular computational setup in Sec.~\ref{sec:setup}.
We move on to study the GRMHD model and emergent radiation in Sec.~\ref{sec:res}, comparing these results with the observations presented above, including non-radiative measurements (e.g., spin-down torques).
We conclude in Sec.~\ref{sec:ccl}.
\\

\section{Methods}
\label{sec:methods}

Our approach consists of two steps: a time-dependent numerical simulation of a turbulent accretion flow interacting with a rotating NS magnetosphere, and a post-processing step to compute the emergent radiation spectrum.

As the first step, we solve the axisymmetric two-di\-men\-si\-onal equations of GRMHD with a modified version of the finite-volume code \textsc{harmpi} \citep{2017MNRAS.467.3604R,2019ascl.soft12014T}, based on 
\textsc{harm2d}
(\citealt{gammie_harm_2003}, \citealt{noble_primitive_2006}),

\begin{equation}
\beal
\nabla_\mu (\rho u^\mu) &= 0 ,\\
\nabla_\mu T_\nu^\mu &= 0, \\
\nabla_\mu \hbox{}^*\!F^{\mu \nu} &= 0,
\label{eq:eqs}
\eeal
\end{equation}
where $\nabla_\mu$ is the covariant derivative and
\begin{equation}
T^{\mu \nu} = \left(\rho \, \mathrm{c}^2 + \epsilon + p + b^2 \right) u^\mu u^\nu  + \left(p+\frac{b^2}{2} \right) g^{\mu \nu} - b^\mu b^\nu 
\label{eq:tmunu}
\end{equation}
is the total energy-momentum tensor; $\rho$, $\epsilon$ and $p$ are the fluid-frame mass density, internal energy density, and thermal pressure, respectively; $u^\mu$ is the four-velocity of the fluid; $b^\mu$ is the magnetic four-vector and $b^2/2$ is the fluid-frame magnetic pressure.
We absorb the factor of $(4\pi)^{-1/2}$ into the definition of the magnetic field.
Finally, $\hbox{}^*\!F^{\mu \nu} \equiv b^\mu u^\nu - b^\nu u^\mu$ is the dual of the electromagnetic field tensor, and $g_{\mu \nu}$ is the metric tensor (see Sec.~\ref{sec:setup}).
We adopt an ideal gas with an adiabatic index of $\gamma = 4/3$. 

We use the hybrid MHD--force-free approach of \cite{parfrey_general-relativistic_2017} [see Appendix B of \cite{2024ApJ...975...57P} for more details], allowing us to evolve the strongly magnetized NS magnetosphere and weakly magnetized flow together.
For this, we introduce a passive scalar $\mathcal{F}$, which satisfies $\nabla_\mu (\mathcal{F} \rho u^\mu) \, {=}\, 0$: $\mathcal{F}\, {\simeq} \, 1$ in the magnetospheric regions and $\mathcal{F}\, {\simeq} \, 0$ in the accretion flow.
We retain the GRMHD solutions when $\mathcal{F}\, {=} \, 0$ and enforce force-free-like values when $\mathcal{F}\, {=} \, 1$. In between, we perform a smooth interpolation. We employ this hybrid procedure only for $r\, {\le}\, \Rlc$, with $\Rlc$ being the light cylinder radius; at larger radii, we evolve the system using the unmodified GRMHD equations. 
Finally, we gradually reduce the strength of the force-free enforcement as $\Rlc$ is approached from below, ensuring a smooth transition to unmodified evolution.


As the second step, we post-process the GRMHD simulation outputs with the GR radiative-transfer code \textsc{grmonty} \citep{dolence_grmonty_2009}, including synchrotron emission, absorption, and inverse Compton scattering.
We assume a thermal electron distribution and adopt the ``fast-light" approximation, i.e., we assume that photon packets travel instantaneously and neglect the time delay between photons emitted from different regions.
For this first application of \textsc{grmonty} to NS accretion, we assume equal electron and proton physical temperatures for simplicity\footnote{
Coulomb electron-proton thermalization (e.g., \citealt{mahadevan_are_1997}) 
is unlikely at the wind--disk interface where electrons are found to be relativistic. This aspect will be improved in future work.
}, so $T_\mathrm{e} \, {\equiv} \, p/(2 \rho)$, where we have assumed that the plasma is comprised purely of ionized hydrogen.
As is usual in the post-processing of GRMHD models (e.g., \citealt{moscibrodzka_radiative_2009}, \citealt{moscibrodzka_revision_2025}), we ignore the emission coming from very high magnetization [$\sigma\, {\equiv}\, b^2 / (\rho \mathrm{c^2}+p+\epsilon)\, {>}\, 20$], near-vacuum regions, because there the density and internal energy are heavily influenced by their background or ``floor'' distributions.
The density floors we adopt are set so that $b^2/\rho \mathrm{c}^2 \, {\leq} \, 10^{4.75}$ inside $\Rlc$ and $b^2/\rho \mathrm{c}^2 \, {\leq} \, 200$ outside $\Rlc$.

\section{Physical and numerical setup}
\label{sec:setup}

We employ a spherical polar grid in $(r,\theta,\phi)$ coordinates and initialize our GRMHD simulations with an axisymmetric torus in hydrodynamic equilibrium \citep{1985ApJ...288....1C,de_villiers_magnetically_2003}.
The torus has its inner radius at $r_\mathrm{in}=60 \, r_\mathrm{g}$, 
its outer radius at $r_\mathrm{out}\approx135\, r_\mathrm{g}$,
and the pressure maximum at $r_\mathrm{max} = 85 \, r_\mathrm{g}$, where $r_\mathrm{g} \equiv \mathrm{G} M_\star/ \mathrm{c}^2$ is the NS gravitational radius and $M_\star$ the NS mass. We choose the torus specific angular momentum, $l \, {\equiv} \, - u_\phi / u_t$, to scale as $\varpropto r^{1/3}$.
This condition is identical to that used in \cite{parfrey_general-relativistic_2017}.

Outside the torus, we set the NS magnetic field via the vector potential that would produce a static dipole of magnetic moment $\mu$ in the limit of a zero-spin (Schwarzschild) spacetime \citep{wasserman_masses_1983}.
We set $\mu \, {=} \, 240$, in units of $4\pi r_\mathrm{g}^3 \sqrt{\rho_0 \mathrm{c^2}}$
; without loss of generality we set $\rho_0 \, {\equiv} \, 1$ in code units.
We deform the stellar magnetic field lines so they flow around the torus but do not enter it \citep[for details see][]{2024ApJ...975...57P}:
 once the simulation starts, the star's rotation immediately opens up these field lines.

Inside the torus, we insert a single poloidal magnetic field loop (i.e., with the magnetic field in the $r$-$\theta$ plane) of covariant magnetic vector potential, $A_\phi \varpropto r^4 \rho^2$, designed to increase the loop magnetic flux by shifting the center of the loop past the pressure maximum of the torus.
We normalize the torus magnetic field strength so that $\max( p)/\max(b^2/2)=100$. 
We seed the magneto-rotational instability (MRI, \citealt{balbus_powerful_1991}) by adding $2\%-$amplitude random perturbations to the gas pressure. 
Furthermore, we choose the torus magnetic field direction near the torus inner edge to be initially anti-parallel to the stellar dipolar field at the equator, thereby ensuring that reconnection between the two is possible \citep{parfrey_general-relativistic_2017}.

Our simulation duration, $t_\mathrm{sim} \, {=} \, 40,\!000 \, r_\mathrm{g}/\mathrm{c}  \, {\sim}\,0.35$~s, is still much shorter than the mode-transition timescale (${\sim}10$~s, \citealt{bogdanov_coordinated_2015}, \citealt{baglio_matter_2023})
and mode duration (hours, e.g., \citealt{linares_neutron_2014}). 
Therefore, we attempt to reproduce low and high modes separately.
For this, we explore two values of the maximum density in the torus: $\rho_\mathrm{max}\, {=} \, \rho_0$ and $\rho_\mathrm{max} \, {=} \, 0.36\, \rho_0$, corresponding physically to larger and smaller large-scale mass supply, respectively.
Indeed, our motivation is to vary the relativistic disk ram pressure ${\sim}\rho \mathrm{c}^2 u^\mu u^\nu$ and therefore explore two disk regimes (e.g., \citealt{miller_magnetohydrodynamic_1997}) around the same NS.
We denote the higher-density torus simulation as the \highmdot{} model and the lower-density torus simulation as the \lowmdot{} model.

We employ a grid with resolution $N_r \times N_\theta \times N_\phi = 1024 \times 512 \times 1$ in the $r$-, $\theta$-, and $\phi$-directions, respectively, extending out to $R_\mathrm{out} = 3,\!000 \, r_\mathrm{g}$.
The grid is logarithmically spaced for $r\, {\le} \, 120\, r_\mathrm{g}$ and is hyper-exponential beyond this radius.
In the $\theta$-direction, the grid is concentrated on the equatorial plane, resulting in an effective $\theta$-cell extent of $0.18$ degrees near the equator and $0.5$ degrees near the poles at $r=\Rlc$. 

Throughout, we set the NS parameters to match tMSP ca\-no\-ni\-cal values.
We adopt the NS mass of $M_\star \, {=} \,1.8\, \mathrm{M_\odot}$ \citep{linares_super-massive_2020}, the NS radius of $R_\star \, {=} \,4\,  r_\mathrm{g} \, {=} \,10.8$~km \citep{ozel_masses_2016}, and the angular velocity of $\Omega_\star \, {=} \,0.03\,  \mathrm{c}/r_{\rm g}$. This results in NS spin period, $P_\star \, {=} \,1.88$~ms, and light cylinder radius, $\Rlc \, {\equiv} \,\mathrm{c}/\Omega_\star \, {\simeq} \, 33.3\, r_\mathrm{g}\, {\simeq}\, 89$~km. 
As a result, in the Kerr-Schild foliation of the Kerr spacetime we use, we set the dimensionless spin parameter, $a_\star  \, {\equiv} \, I_\star \Omega_\star \mathrm{c} / \mathrm{G} M_\star^2 \, {=} \,0.2$, with $I_\star$ being the star's momentum of inertia \citep{ravenhall_neutron_1994}.
The corotation radius, at which the stellar angular velocity equals the Keplerian orbital angular velocity, is $R_\mathrm{co} \, {\equiv} \, (\Rlc - a_\star r_\mathrm{g})^{2/3} r_\mathrm{g}^{1/3}\, {\simeq}\, 10.3 \, r_\mathrm{g} \, {\simeq}\, 28$~km. 
In the radiative transfer calculation, we adopt a distance to the NS of $1.37$~kpc to match that of PSR J1023+0038 \citep{deller_parallax_2012}.
For this proof-of-concept study,  rather than adopting a specific inclination angle, we consider spectral energy distributions (SEDs) averaged over the entire, $4\pi$, solid angle. 
We post-process the GRMHD data every $400 \, r_\mathrm{g}/\mathrm{c}$ with \textsc{grmonty} using $10^4$ superphotons, which we accumulate at a large radius and compute the flux density $F_\nu$ from the radio to gamma-ray frequencies. 
To ensure higher quality of the spectrum, the superphoton weights are based on the local synchrotron emissivity function (averaged over the entire solid angle) rather than being uniform across frequency bins \citep{dolence_grmonty_2009}.

In non-radiative GRMHD simulations, the density (or, equivalently, energy) normalization is scale-free.
However, this normalization enters the synchrotron frequency and emissivity function, as well as the absorption and scattering opacities.
Thus, in \textsc{grmonty} we set the gas density normalization, 
$\rho_0 \, {=} \, 1.2 \times 10^{-9}\, \mathrm{g \, cm^{-3}}$,
so that the \lowmdot{} model produces an X-ray luminosity equal to that of the observed high mode, $\Lx\,  {\sim}\,  10^{33}\, \mathrm{erg \, s^{-1}}$ (e.g., \citealt{linares_x-ray_2014}, \citealt{archibald_accretion-powered_2015}). 
We then use the same density normalization to post-process the \highmdot{} model.
This density normalization also sets the scale for the magnetic field strength, allowing us to check a posteriori for consistency between the corresponding NS magnetic field and observational values (see Sec.~\ref{sec:NS_mag_field}). 



\section{Results}
\label{sec:res}

\subsection{Time evolution and modes} 
\label{sec:res_1}

\begin{figure*}
\fig{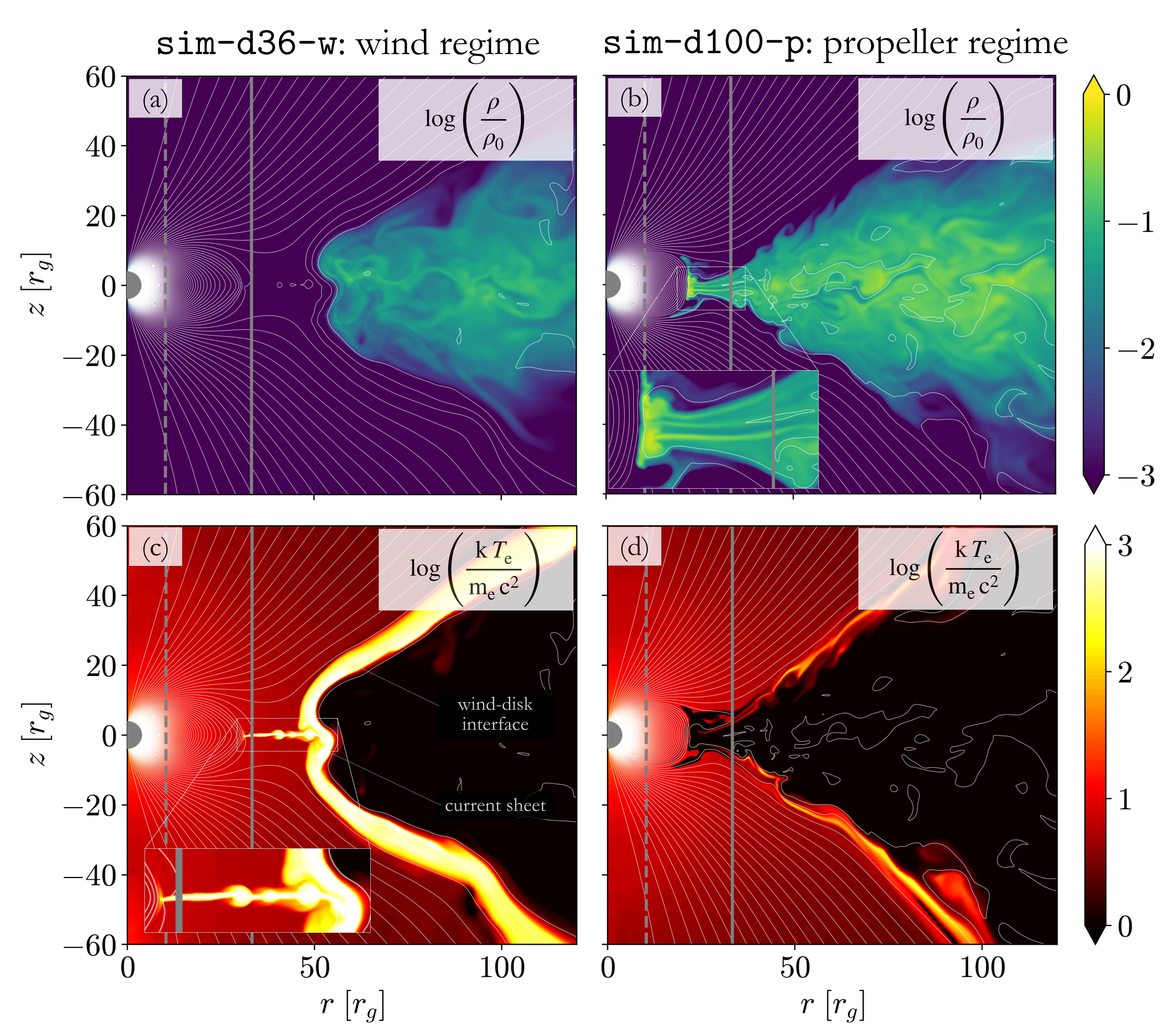}{\textwidth}{}
\caption{
Our simulated slices of density (panels a,b) and temperature (panels c,d) reveal that the wind--disk interface is hotter when the disk is located beyond the light cylinder: this predominantly occurs in the model with lower inflow rate, \lowmdot{}, as seen in panel (d).
We show the system at $t \, {=} \, 17,520 \, r_{\rm g} / {\rm c}$ in the \lowmdot{} model (left panels) and at $t \, {=} \, 20,600\, r_{\rm g} / {\rm c}$ in the \highmdot{} model (right panels).
\textbf{(panel a):} \lowmdot{} model in the {\lq}wind{\rq} regime, with the accretion flow maintained outside $\Rlc$.
\textbf{(panel b):} \highmdot{} model in the propeller regime, with the accretion flow between $R_\mathrm{co}$ and $\Rlc$. 
\textbf{(panel c):} Comparatively higher temperature in the current sheet and pulsar wind--disk interface in the wind regime (from the \lowmdot{} model here).
\textbf{(panel d):} Temperature distribution in the propeller regime (from the \highmdot{} model here).
The pulsar wind--disk collision heats up the plasma to relativistic energies and leads to synchrotron-powered X-ray emission (see Fig.~\ref{fig:SED_low_high}).
Poloidal magnetic field lines are overplotted in white.
Vertical lines indicate $R_\mathrm{co}$ (dashed) and $\Rlc$ (solid line).
Animated versions of panels (a) to (d) can be found here: \href{https://nuage.osupytheas.fr/s/FLnjjbEXLLXLmNa}{https://nuage.osupytheas.fr/s/FLnjjbEXLLXLmNa}, \href{https://nuage.osupytheas.fr/s/3Fw94Bjde9aiSLD}{https://nuage.osupytheas.fr/s/3Fw94Bjde9aiSLD},
\href{https://nuage.osupytheas.fr/s/gNKGnND8fCkMgiT}{https://nuage.osupytheas.fr/s/gNKGnND8fCkMgiT}, \href{https://nuage.osupytheas.fr/s/md9CBoTBrGBA66W}{https://nuage.osupytheas.fr/s/md9CBoTBrGBA66W}.}
\label{fig:rho_temp_low_high}
\end{figure*}

We first focus on the flow dynamical regime reached in our two GRMHD models before discussing the time dependence.
Figure~\ref{fig:rho_temp_low_high}(a,b) shows the representative density maps, with magnetic field lines overplotted, for each of our models. 
Figure~\ref{fig:rho_temp_low_high}(a) shows that in the \lowmdot{} model the pulsar wind prevents the accretion flow from entering the light cylinder altogether. So, the flow remains mostly outside $\Rlc$, and the last closed field line stays around $\Rlc$; we refer to this regime as \lowmdot{}\texttt{-w} for {\lq}wind{\rq}. 
Meanwhile, Fig.~\ref{fig:rho_temp_low_high}(b) shows that in the \highmdot{} model, the flow penetrates the NS magnetosphere and opens the magnetospheric field lines, which stop the flow at the magnetospheric radius, $\Rm$, which is located between the corotation, $R_\mathrm{c}$, and light cylinder, $\Rlc$, radii (shown in Fig.~\ref{fig:rho_temp_low_high} with the dashed and solid gray vertical lines, respectively).
Animations (see Fig.~\ref{fig:rho_temp_low_high} caption) reveal that
portions of the gas residing in the magnetospheric region ($r<\Rlc$) are expelled along the disk surface, consistent with a propeller-like regime; we refer to this regime as \highmdot{}\texttt{-p} for {\lq}propeller{\rq}.

The above behavior is as expected: as the disk ram pressure decreases, the magnetosphere stops gas inflow at
$\Rm$.
However, interestingly, continuous changes in $\rho_\mathrm{max}$ result in two distinct disk regimes, i.e., the system exhibits bimodality.
Specifically, the simulations with the higher density than the \highmdot{} model, $\rho_0\lesssim\rho_\mathrm{max} \lesssim 2.25\rho_0$, result in qualitatively similar values of the magnetospheric radius. 
In contrast, the simulations with even lower densities than the \lowmdot{} model, $\rho_\mathrm{max} \, {\lesssim} \, 0.36\, \rho_0$, 
result in qualitatively similar exclusion of gas from the light cylinder.
Within each of these distinct categories, $\Rm$ is a smoother function of $\rho_\mathrm{max}$ than across these categories, with larger $\rho_\mathrm{max}$ resulting in smaller $\Rm$.

Figure~\ref{fig:rho_temp_low_high}(c) shows that \lowmdot{}\texttt{-w} develops two high 
temperature regions: (i)~the equatorial current sheet, seen in Fig.~\ref{fig:rho_temp_low_high}(c) inset, which forms plasmoids (suggesting the current sheet is in the fast reconnection regime, \citealt{2010PhRvL.105w5002U}), and (ii)~the interface between the wind and disk surface
(with magnetization up to $\sigma \sim 10$). 
Figure~\ref{fig:rho_temp_low_high}(d) shows that both regions are absent in \highmdot{}\texttt{-p}, although the disk surface is slightly hotter than the disk plane.
To quantify the pulsar wind power dissipation, which is presumably responsible for this heating, we compute the radial profiles of all luminosity components: total, rest-mass, electromagnetic and thermal luminosities.
Outside the light cylinder, $\Rm \, {>} \, \Rlc$, we indeed find a higher thermal luminosity in \lowmdot{}\texttt{-w} than for disk-free pulsars.
We attribute this difference to the wind luminosity intercepted by the disk; see Appendix~\ref{app:dissipation} for details.

\begin{figure*}[!htbp]
\gridline{
\includegraphics[width=0.5\textwidth]{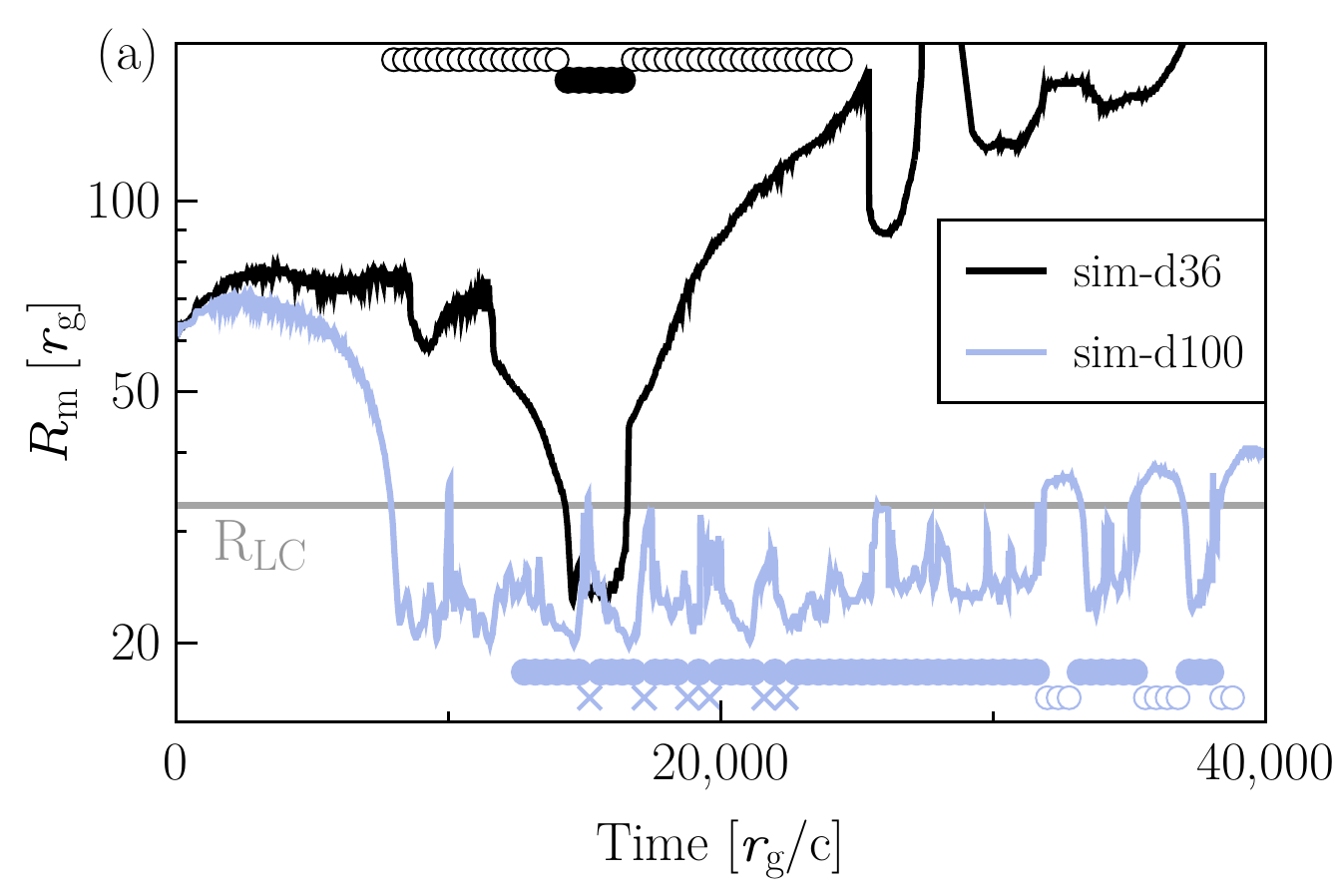}
\includegraphics[width=0.5\textwidth]{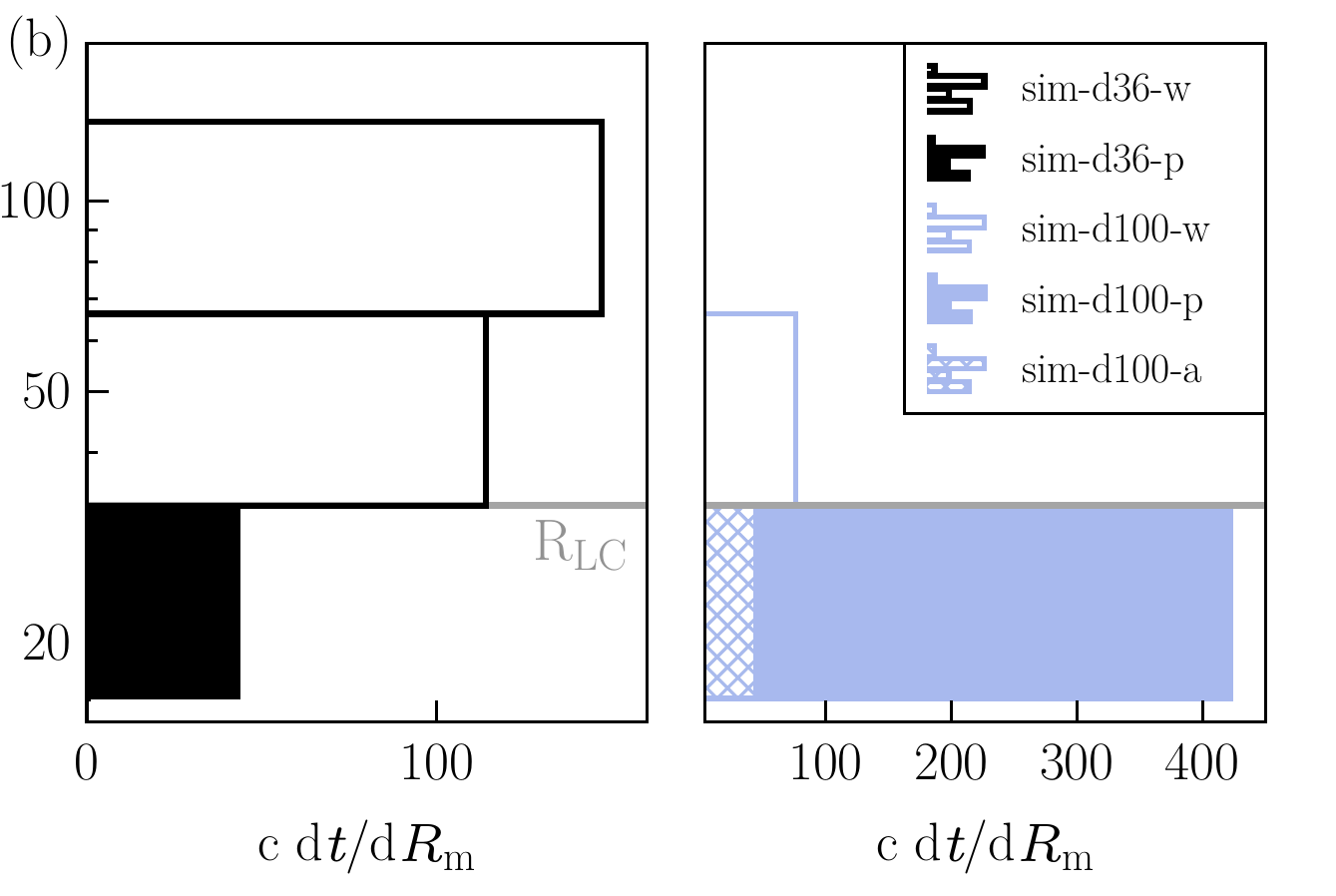}}
\caption{
The time evolution (panel a) and histogram (panel b) of the magnetospheric radius, $\Rm$, shows bimodality with respect to the light cylinder radius, $\Rlc$, and the main accretion regime depends on the initial torus density.
\textbf{(panel a):} In the \highmdot{} model (light blue) $\Rm$ lies mainly between $R_\mathrm{co} \, {\simeq} \, 10.3 \, r_\mathrm{g}$ (located outside the plotted range) and $\Rlc \, {\simeq} \, 33.3\, r_\mathrm{g}$.
In the \lowmdot{} model (black), the pulsar wind pushes the flow out of $\Rlc$ most of the time.
\textbf{(panel b):} Histogram showing the time spent for each binned value of $\Rm$ and its link to the accretion regime denoted {\rq}w{\lq} for {\lq}wind{\rq}, {\rq}p{\lq} for {\lq}propeller{\rq} and {\rq}a{\lq} for transient accretion columns (or simply {\lq}accretor{\rq}), among the epochs covered by our \textsc{grmonty} post-processing step.
The instantaneous times of each post-processed GRMHD snapshot are indicated by the small boxes on panel (a), where we follow the same legend as in panel (b): filled symbols for propeller, empty symbols for wind and crosses for accretor.
} 
\label{fig:rdisk}
\end{figure*}

We compute $\Rm$ by finding the smallest radius at which the equatorial density (within $7.5$~deg of the equator) exceeds 1\% of the initial maximum density ($\rho>\rho_\mathrm{max}/100$).
Figure~\ref{fig:rdisk}(a) shows the time evolution of $\Rm$ for both of our models.
The black and light blue curves show that at the beginning of both the \lowmdot{} and \highmdot{} models, respectively, the magnetospheric radius starts out at $\Rm=60\, r_\mathrm{g}$. This makes sense, because the magnetosphere initially fills the entire space inside the inner radius of the initial torus (Sec.~\ref{sec:setup}). As each of the simulations starts, the pulsar wind launches on the timescale of the NS rotational period and pushes the inner edge of the torus out of the way: this causes $\Rm$ to initially increase in both simulations. However, as the MRI-driven turbulence develops in the torus, it viscously expands, falls inward, and pushes back on the pulsar wind: this causes $\Rm$ to decrease.  
The viscous expansion and inflow happen roughly on the Keplerian timescale at the magnetospheric radius, $t_{\rm m}\sim 2\pi \times (\Rm/r_{\rm g})^{3/2}r_{\rm g}/\mathrm{c}\sim3,\!000\, r_{\rm g}/{\rm c}$. Over this timescale, the higher ram pressure of the inflowing gas in the \highmdot{} model slows down the pulsar wind expansion more strongly than in the \lowmdot{} model, and the blue curve undershoots the black curve. The higher-density gas in the \highmdot{} model takes time to generate enough ram pressure to crush the wind and cause the blue $\Rm$ curve to drop precipitously around $t\simeq8,\!000\, r_{\rm g}/{\rm c}$. The lower-density gas in the \lowmdot{} model takes even longer to do the same, and the black curve eventually also drops below $R_{\rm LC}$ at $t\simeq15,\!000\, r_{\rm g}/{\rm c}$. However, after the gas squeezes the magnetosphere, the evolution of the two models qualitatively diverges, and they spend most of their time in one of the distinct wind or propeller regimes.

Figure~\ref{fig:rdisk}(b) shows the histogram quantifying how much time the system spends 
at a given value of $\Rm$.
Together with Figure~\ref{fig:rdisk}(a), this shows that the \highmdot{} model spends most of the time in the classical propeller regime: the innermost flow resides mostly between the corotation, $R_\mathrm{co}\, {\simeq} \, 10.3 \, r_\mathrm{g}$, and light cylinder, $\Rlc\, \, {\simeq} \, 33.3\, r_\mathrm{g}$, radii.
This model also features transient episodes of magnetically channeled accretion onto the stellar surface (labelled \highmdot{}\texttt{-a} for {\lq}accretor{\rq}), when the flow slides down the field lines off the equatorial plane (Fig.~\ref{fig:rho_temp_low_high}(b)), or expulsion from the light cylinder (labelled \highmdot{}\texttt{-w}). 
Meanwhile, after the flow enters the light cylinder in the \lowmdot{} model at $t \, {\simeq}\, 15,\!000 \, r_\mathrm{g}/\mathrm{c}$, the pulsar wind rapidly expels it, so that we have $\Rm \, {>} \, \Rlc$  most of the time.
Similarly, the occurrences of $\Rm \, {\gtrsim} \, \Rlc$ become more frequent at the end of the \highmdot{} simulation as the mass reservoir is depleted. 
Overall, this reveals the aforementioned bimodality in the behavior of $\Rm$ relative to $\Rlc$: generally, the \highmdot{} model exhibits $\Rm \, {<} \, \Rlc$, and is predominantly in the propeller regime, and the \lowmdot{} model exhibits $\Rm \, {>} \, \Rlc$, and is predominantly in the wind regime.

\begin{figure}
\centering
\includegraphics[width=\minof{0.75\textwidth}{\columnwidth},trim=0.25cm 0 1cm 0,clip]{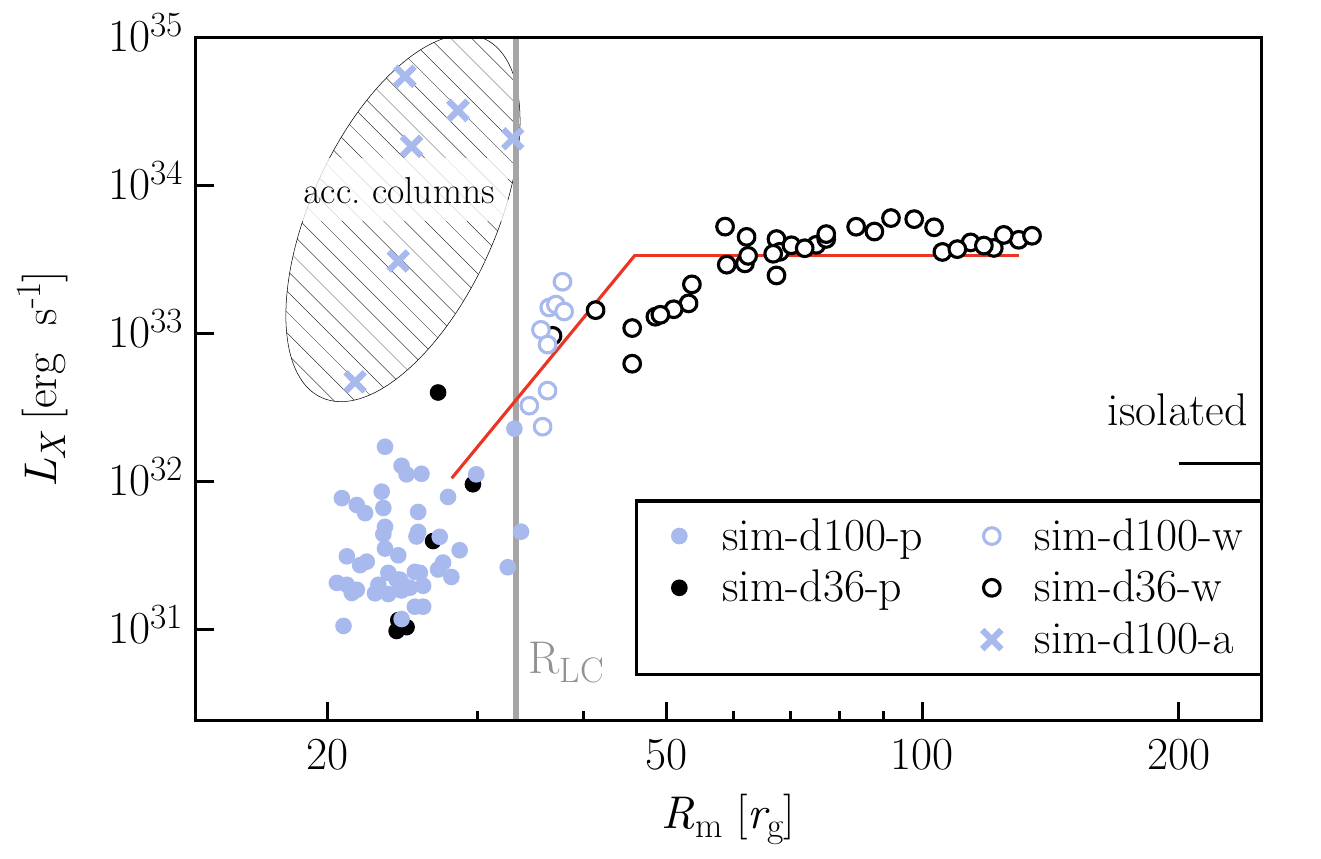}
\caption{
Simulated X-ray luminosity, $\Lx$, increases monotonically with increasing magnetospheric radius, $\Rm$, and saturates at $\Rm\gtrsim \Rlc$.
We attribute this to the wind luminosity intercepted by the disk and the wind dissipated fraction both increasing with increasing $\Rm$ and saturating once $\Rm$ is outside the light cylinder (Appendix~\ref{app:dissipation}). 
Data from the \highmdot{} model are shown in light blue, data from the \lowmdot{} model are shown in black, and the average X-ray luminosity of the isolated pulsar model is indicated with the short horizontal line.
Symbols follow the same legend as in Fig.~\ref{fig:rdisk}.
The red curve is plotted to guide the eye across the correlation.
The few accretion column events generally overshoot the observed X-ray luminosity (see the last paragraph of Sec.~\ref{sec:res_1}). 
}
\label{fig:L_X_Rm}
\end{figure}

We consider the following time intervals for the analysis of the SEDs: $8,000\, r_\mathrm{g}/\mathrm{c}\le t\le 24,000  \, r_\mathrm{g}/\mathrm{c} \sim 0.14$~s for \lowmdot{}, before the disk is dispersed by the wind, and $12,800 \, r_\mathrm{g}/\mathrm{c} \le t\le 38,800 \, r_\mathrm{g}/\mathrm{c} \sim 0.23$~s for \highmdot{}. 
Since we find extra dissipation from wind--disk interaction when $\Rm \, {>} \, \Rlc$ (Appendix~\ref{app:dissipation}) with a correspondingly higher temperature at the wind--disk intersection (Fig.~\ref{fig:rho_temp_low_high} (a,c)), and obtain a bimodality in $\Rm$ (Fig.~\ref{fig:rdisk}), we expect also a bimodality in X-ray luminosity with higher luminosity when $\Rm \, {>} \, \Rlc$.


Indeed, Figure~\ref{fig:L_X_Rm} shows this positive correlation between the instantaneous value of $\Rm$ and
intrinsic source X-ray luminosity, $\Lx$ (which we compute by integrating $F_\nu$ from $0.5$ to $10$~keV).
Indeed, for $\Rm \, {\lesssim} \, \Rlc$, $\Lx$ is smaller than for $\Rm \, {>} \, 2 \,  \Rlc$ because of the absence of wind-power dissipation.
This configuration could correspond, at least qualitatively, to the low mode, or to a more extreme version of it, because it features a smaller flux than observations of the mode 
\citep[$\Lx \, {\sim} \, 6-9\, {\times} \, 10^{32} \, \mathrm{erg \, s^{-1}}$;][]{linares_x-ray_2014}. Meanwhile, $\Lx$ increases with 
increasing $\Rm$.
The $\Rm \, {>} \, \Rlc$ configuration, i.e. the wind regime, is compatible with the high mode, because $\Lx$ is similar to the observed values for the mode (${\gtrsim}\,  10^{33}\, \mathrm{erg \, s^{-1}}$; see also Sec.~\ref{sec:setup}).
This implies that $\Rm/\Rlc$ is the key parameter controlling the X-ray luminosity.

Figure~\ref{fig:L_X_Rm} also shows that for $\Rm/\Rlc \, {\gtrsim} \, 2 $, $\Lx$ saturates at a constant value that exceeds that for the isolated pulsar, in which case the hot plasma is concentrated in the current sheet. 
We interpret this saturation as the pulsar wind power intercepted by the disk remaining stable. 
Such stability is likely favored by the wind power being concentrated near the equatorial plane, which also coincides with the disk plane in our model.
Hence, the aforementioned saturation can explain the stability of the high mode flux. 
Alternatively, the high mode stability could be tied to the stability of the disk-intercepted fraction of the pulsar wind luminosity, which is set by the solid angle subtended by the disk and is independent of $\Rm$ for uniform aspect-ratio disks.
For $\Rm \, {<} \, \Rlc$, the dispersion we observe in $\Lx$ suggests that the low mode might not be as stable as the high mode; however, we cannot conclude on this yet because our models in the propeller regime (filled circles) undershoot the observed flux values.

The only exceptions to low $\Lx$ for $\Rm \, {\leq} \, \Rlc$ are concurrent with the \highmdot{}\texttt{-a} regime.
During most such events of accretion through columns, 
the flow reaches the NS surface.
Among the \highmdot{}\texttt{-a} points, the lowest $\Lx$ point corresponds to the flow not reaching the NS surface.
From the measured X-ray flux, the SED, and the constraints on the torques, we conclude that accretion column events
are rare or absent in the observations (see Appendices~\ref{app:mode_combi} and \ref{app:torques}). 

Finally, because each model produces a paucity of points at $\Rm\simeq \Rlc$ (also visible in Fig.~\ref{fig:rdisk}(a)), the bimodality in $\Rm$ carries over to the X-ray luminosity.
Henceforth, we associate the high-$\Lx$ mode with our \lowmdot{}\texttt{-w} model shown in Fig.~\ref{fig:rho_temp_low_high} (left): the wind regime with the inner disk outside the light cylinder ($\Rm \, {>} \, \Rlc$).
Similarly, we associate most of the low-$\Lx$ mode with the \highmdot{}\texttt{-p} model shown in Fig.~\ref{fig:rho_temp_low_high} (right): a propeller-like regime with the inner disk inside the light cylinder ($\Rm \, {<} \, \Rlc$ and no accretion columns).

\subsection{Multi-wavelength spectral energy distribution}
\label{sec:res_2}


\begin{figure*}
\begin{center}
\includegraphics[height=0.35\textwidth,trim=0 0 1cm 0,clip]{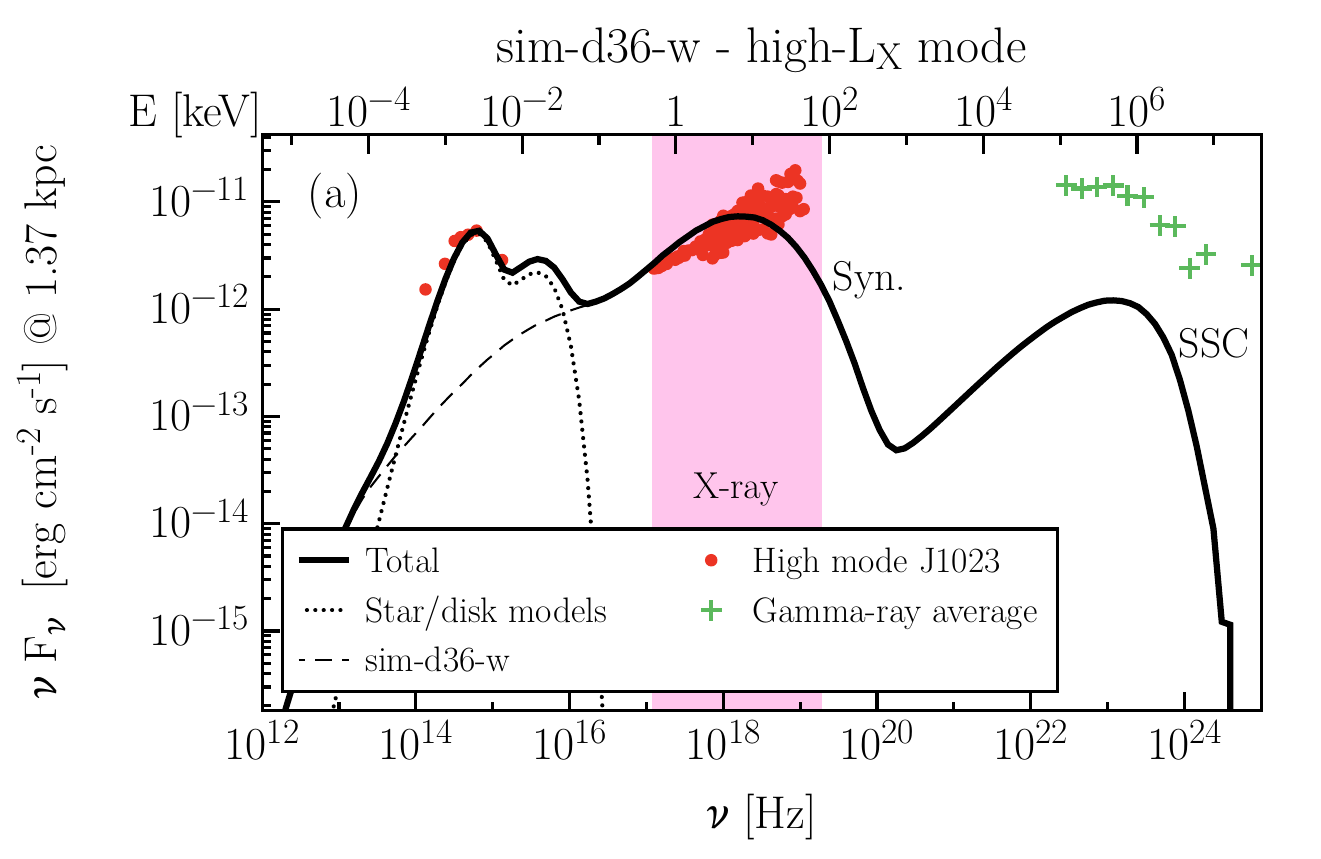}
\includegraphics[height=0.35\textwidth,trim=2.2cm 0 1cm 0,clip]{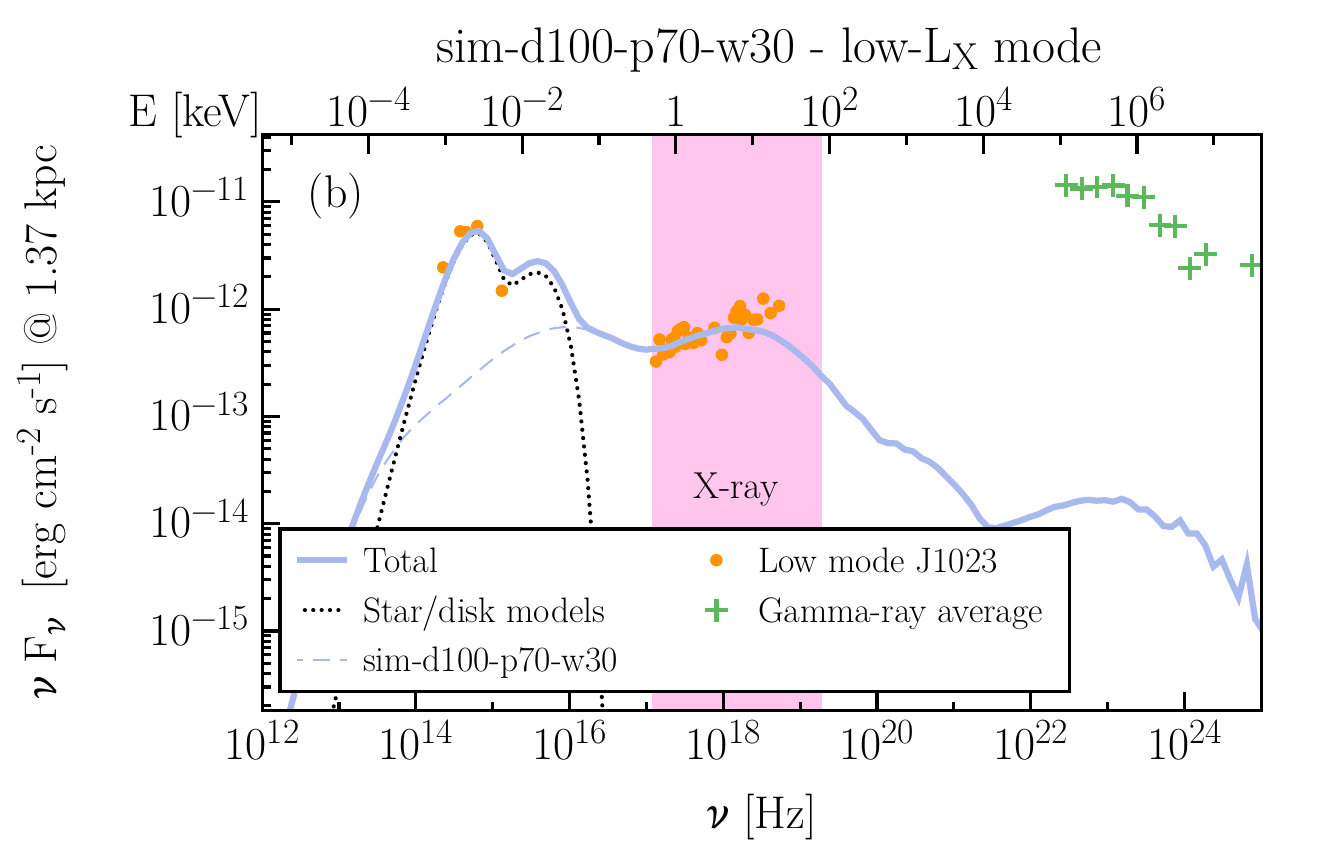}
\end{center}
\vspace{-0.7cm}
\caption{Composite SEDs of our models show good agreement with the multiwavelength spectra of low- and high-$\Lx$ modes.
Optical to gamma-ray spectral energy distributions from \lowmdot{}\texttt{-w} corresponding to $100\%$ wind regime in the lower-density run, which we attribute to the high mode  (\textbf{panel a}) and from \highmdot{}\texttt{-p70-w30} corresponding to $70\%$ propeller and $30\%$ wind regimes in the higher-density run, which we attribute to the low mode (\textbf{panel b}).
Contributions include the time-averaged GRMHD (dashed), optical companion, and thermal thin-disk models (dotted).
We fixed the density normalization to fit the high mode 0.5--10~keV flux (Sec.~\ref{sec:setup}).
For the \lowmdot{}\texttt{-w} model, we indicate the dominant emission mechanism contributing to different parts of the SEDs, synchrotron and synchrotron self-Compton.
X-ray (orange/red dots) and gamma-ray (green pluses; averaged over both modes) data for PSR J1023+0038 are taken from \citealt{baglio_matter_2023} and \citealt{papitto_pulsating_2019}, respectively, and plotted for comparison.
The X-ray band (0.5-80~keV) is shown in pink.
}
\label{fig:SED_low_high}
\end{figure*}

For comparison with optical and UV data, we need to account for (i) the emission from the companion star and (ii)~the companion-fed thermal, 
optically thick and geometrically thin disk (non-irradiated for simplicity, unlike \citealt{baglio_matter_2023}) whose interaction with the NS dipole at much smaller distances we simulate in this work. We use analytic models for both of the components.
Similar to \citet{baglio_matter_2023}, for the optical emission from the companion, we assume blackbody emission from a star of radius $R_\mathrm{c} \, {=} \, 0.5 \, R_\odot$ and temperature $T_\mathrm{c}\, {=} \, 6388$~K; we also adopt the thin disk model of \cite{shakura_black_1973} with radiative efficiency $\eta_\mathrm{eff}\, {=} \, 0.1$ and accretion rate $\dot{M} \, {=} \, 3\times 10^{-4} \, \dot{M}_\mathrm{Edd}$, where $\dot{M}_\mathrm{Edd}=2.5 \times 10^{18} ~\mathrm{g \, s^{-1}}$ is the Eddington accretion rate for an $M_\star \, {=} \,1.8\, \mathrm{M_\odot}$ NS.
We set the inner and outer thin disk radii to $10^3 \, r_\mathrm{g} \, {=} \,2.7 \times 10^8$~cm and $10^5 \, r_\mathrm{g} \, {=} \,2.7\times 10^{10}$~cm, respectively. 
This represents an outer optically thick accretion disk that bridges our GRMHD simulations of the inner optically thin flow (initially extending out to ${\sim}\, 10^2 \, r_\mathrm{g}$, Sec.~\ref{sec:methods}) and the circularization radius (at ${\sim}\, 10^5 \, r_\mathrm{g}$ for J1023, see e.g., Eq.~4.20 of \citealt{frank_accretion_2002}).

Figure~\ref{fig:SED_low_high} shows the SED obtained from the joint GRMHD and radiative transfer models (Sec.~\ref{sec:methods}), including the contributions from the analytical models for the companion star and outer disk (full lines) and showing them separately (dotted line). 
We overplot the observational data of J1023 with symbols, taken from \cite{baglio_matter_2023} and \cite{papitto_pulsating_2019}.
We show these for the \lowmdot{}\texttt{-w} on panel (a), that represents $100\%$ wind regime. 
As we discuss above, we associate the \lowmdot{}\texttt{-w} regime with the high-$\Lx$ mode and obtain the GRMHD contribution to the spectra in Fig.~\ref{fig:SED_low_high} by
averaging $F_\nu$ over the corresponding epoch of the \lowmdot{} simulation. 
For the low-$\Lx$ mode, we first computed the time-average $F_\nu$ of the \highmdot{}\texttt{-p} model but found it to undershoot the observed flux (Appendix~\ref{app:mode_combi}).
However, note that although our \highmdot{} simulation featured a higher density normalization, it did not end up spending $100$\% of the time in the propeller regime and ended up switching back and forth between the propeller and wind regimes several times.
What if the same switching takes place in nature in the low-\Lx{} mode?
Indeed, because the observational data span much longer timescales than our simulations, it is plausible that the low-$\Lx$ mode is not a pure propeller regime and features contamination from the wind regime.
To model this, we compute a weighted average between the SEDs corresponding to each of these regimes (see Appendix~\ref{app:mode_combi}); the weight reflects the fraction of the observational exposure time spent in a given regime.
Figure~\ref{fig:SED_low_high}(b) shows \highmdot{}\texttt{-p70-w30}
a weighted combination of $70$\% propeller and $30$\% wind--disk heating epochs in the \highmdot{} model (light blue curves). 

Pink bands in Fig.~\ref{fig:SED_low_high} highlight the X-ray band and the agreement of the \lowmdot{}\texttt{-w} model, shown with black line, with the observed high mode, shown with red dots.
To obtain such an agreement, we have chosen the gas density normalization appropriately (Sec.~\ref{sec:setup}).
Our model naturally produces a slope or spectral index $\alpha\, {=}\, 0.4$ (for $\nu F_\nu \, {\propto}\, \nu^{\alpha}$), similar to the observed X-ray data \citep[which show a hard X-ray spectrum with photon index $\Gamma=2-\alpha \simeq 1$--$2$;][]{linares_x-ray_2014}.
However, our cut-off frequency $\nu \, {\sim}\, 10^{18}$~Hz is smaller than the observations \citep[][found no spectral cut-off up to 79 keV]{tendulkar_nustar_2014}.
On the other hand, 
the \highmdot{}\texttt{-p70-w30} model reaches reasonable agreement, in terms of flux and spectral slope, with the low mode observed data.
This suggests residual pulsar wind--disk heating even in the low mode.

The GRMHD models of the inner flow contribute from the optical to the X-ray bands and up to the gamma-ray band, with negligible radio flux.
The optical band is dominated by the contribution from the companion, but the GRMHD models contribute a few percent of the total flux, in agreement with \cite{baglio_polarized_2025}.
Similarly, in the UV band the GRMHD model component is small compared to the thermal disk one.
The gamma-ray flux is negligible 
($\nu F_\nu \, {\sim} \, 10^{-14} \, \mathrm{erg \, cm^{-2} \, s^{-1}}$) in the \highmdot{}\texttt{-p70-w30} model but it is orders of magnitude higher for the \lowmdot{}\texttt{-w} model, at $\nu F_\nu \, {\sim}\, 10^{-12} \, \mathrm{erg \, cm^{-2} \, s^{-1}}$, reaching ${\sim}\, 0.1$ of the observed level.
This large underestimation of the gamma-ray flux is plausibly due to our assumed thermal distribution of electrons.

By switching off absorption and/or scattering, we identify their contributions to the overall SED; by switching off both, only pure synchrotron emission remains.
We see that, in the wind regime, the X-ray flux (0.5--80~keV) is produced by thermal synchrotron.
The gamma-ray bump (${\sim} \, 10^{-3}$--10~GeV) is due to inverse Compton scattering of the X-ray photons off the relativistic electrons in the same wind--disk interface region, i.e., synchrotron self-Compton (SSC).
On the other hand, in the propeller regime, the UV bump is mainly due to synchrotron emission and the X-ray flux to Compton scattering.
Overall, the emission-only radio flux is $\nu F_\nu \, {\sim}\, 10^{-15} \, \mathrm{erg \, cm^{-2} \, s^{-1}}$, 
but this becomes negligible once synchrotron self-absorption is accounted for in our post-processing (\citealt{dolence_grmonty_2009}; \citealt{papitto_propeller_2014}).
At $\nu \, {\gtrsim} \, \text{few}\times 10^{13}$~Hz absorption plays no role.

\subsection{Density normalization and neutron star magnetic field}
\label{sec:NS_mag_field}
As described in Sec.~\ref{sec:setup}, we have chosen the density normalization to obtain X-ray fluxes from the \lowmdot{} model similar to the observed values in the high mode and applied the same normalization for all models.
Here, we aim to compare the NS dipolar magnetic moment associated with this chosen density normalization to observational constraints; this is an independent test of our model.
We derive the magnetic moment in our GRMHD models,
\begin{equation}
    \mu_\mathrm{cgs} = 10^{26}
    \frac{ \mu }{240} 
    \left( \frac{M_\star}{1.8\,  \mathrm{M_\odot}} \right)^3 
    \left( \frac{ \rho_0 }{1.2 \times 10^{-9} \mathrm{g \, cm^{-3}} }  \right)^{1/2}
\mathrm{G \, cm^3},
\label{eq:mu_cgs}
\end{equation}
where $\rho_0$ is the characteristic density; here we set it to $1.2 \times 10^{-9} \mathrm{g \, cm^{-3}}$ to fit the high mode X-ray flux (see Sec.~\ref{sec:setup}).
By doing so, the \highmdot{} model has an initial, maximum torus density of $\rho_\mathrm{max} \, {=} \, 1.2 \times 10^{-9} \mathrm{g \, cm^{-3}}$ in physical units, while it is $\rho_\mathrm{max} \,  {=} \, 4.3  \times 10^{-10} \mathrm{g \, cm^{-3}}$ in the \lowmdot{} model. 
We recall that $\rho_0$ and $\rho_\mathrm{max}$ are two independent parameters: while we showed that $\rho_\mathrm{max}$ influences the GRMHD dynamical regime (at fixed $\mu$), $\rho_0$ sets the physical units of the density and of the NS magnetic moment.
Our magnetic moment corresponds to the surface magnetic field via $\mu_\mathrm{cgs} \,  {=} \, B_\star R_\star^3$ with $R_\star\, {=} \, 10.8$~km, giving $B_\star \, {=} \, 7.9 \times 10^7$~G.
In comparison, \cite{deller_parallax_2012} deduced the surface magnetic field of $9.6 \times 10^7$~G for J1023. Similar values, of order ${\sim}10^8$~G are estimated for the other tMSPs (\citealt{papitto_transitional_2022} and references therein).
Our numerical models, therefore, agree with the measured NS magnetic fields.

\begin{figure}
\centering
\includegraphics[width=\minof{0.75\textwidth}{\columnwidth},trim=0.25cm 0 1cm 0, clip]{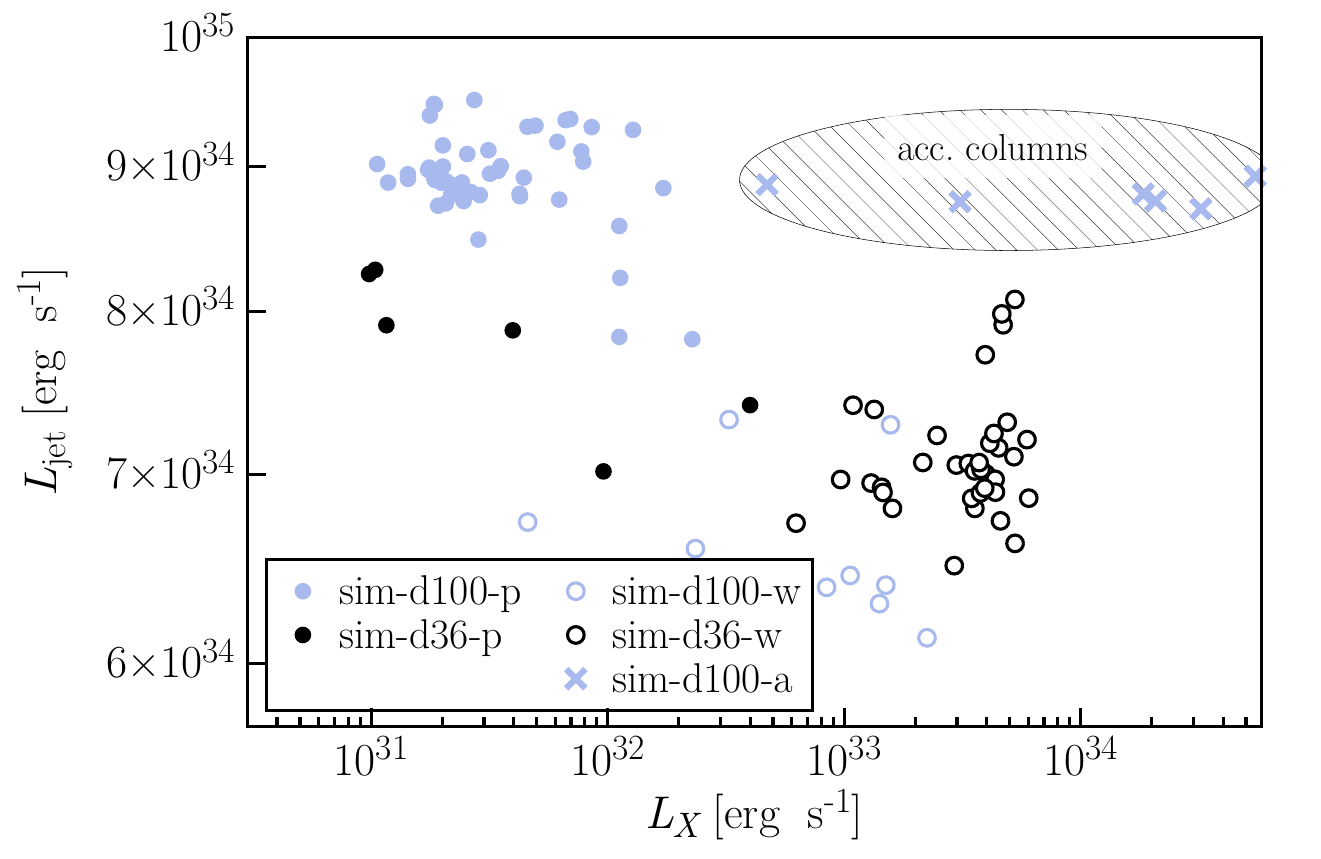}
\caption{Our simulations naturally exhibit an anti-correlation between the jet and X-ray emission. This can explain the observed radio--X-ray anti-correlation in the disk state.
We distinguish data points corresponding to propeller (filled circles), wind (empty circles) and accretor (crosses) regimes.
The anti-correlation is visible within each individual simulation.}
\label{fig:Ljet}
\end{figure}

\subsection{Anti-correlated jet and X-ray variability}
\label{sec:anticor_jet_X}

We integrate the jet electromagnetic luminosity $L_\mathrm{jet} \, {\equiv} \, - \int (b^2 u^r u_t - b^r b_t) \sqrt{-g} \mathrm{d}\theta \mathrm{d}\phi$ over a sphere of radius $\Rlc$ (only counting cells with $\sigma {>} 1$) and
convert it to physical units using the \textsc{grmonty} normalization (Sec.\ref{sec:NS_mag_field}).
For reference, the flat-spacetime wind power for an isolated pulsar is $\mu_\mathrm{cgs}^2 \Omega_\star^4 / \mathrm{c^3} \, {=} \, 4.6 \times 10^{34}$~erg/s, taking $\mu_\mathrm{cgs} \, {=} \, 1.0 \times 10^{26}\, \mathrm{G \, cm^3}$ (Eq.~\ref{eq:mu_cgs}) and $\Omega_\star$ from Sec.~\ref{sec:setup}.
Figure~\ref{fig:Ljet} shows that $L_\mathrm{jet}$ increases by up to $50\%$ when the system transits from the wind regime to the propeller regime (see, e.g., \citealt{parfrey_torque_2016}) because the accretion flow opens the previously closed magnetic flux. 
Similarly, the jet variability within the wind regime results from the regular opening of few field lines at $\Rlc$ in the equatorial plane, leading to magnetic reconnection and plasmoid formation (Fig.~\ref{fig:rho_temp_low_high} (a)).
As seen in Fig.~\ref{fig:Ljet}, we find an anti-correlation between the jet and X-ray emission\footnote{To quantify this association, visible by eye as two anti-correlated clusters of points, with a Fisher statistical test, we assess the association between the two following binary variables: high vs.\ low $L_\mathrm{jet}$ (threshold value: $7.5 \times 10^{34} \, \mathrm{erg \, s^{-1}}$) on the one hand, and high vs.\ low $\Lx$ (threshold value: $2 \times 10^{32} \, \mathrm{erg \, s^{-1}}$) on the other hand. The probability (p-value) of the null hypothesis ``there is no association between the two variables'' is $5 \times 10^{-15}$.}.
As $\Rm$ increases, fewer NS field lines are opened by the accretion flow and $L_\mathrm{jet}$ decreases accordingly.
In the meantime, once $\Rm$ increases above $\Rlc$, pulsar wind heating occurs so $\Lx$ increases, resulting in anti-correlated X-ray and jet emission.
Hence, the high-$\Lx$ mode corresponds to a low value of $L_\mathrm{jet}$, and the low-$\Lx$ mode (considered as $70\%$ of \highmdot{}\texttt{-p} and $30\%$ of \highmdot{}\texttt{-w}) corresponds to a high value of $L_\mathrm{jet}$.
This is consistent with the radio--X-ray anti-correlation observed in the sub-luminous disk state \citep{bogdanov_simultaneous_2018}, if one assumes that the jet powers
the observed radio flux (e.g., \citealt{bogdanov_coordinated_2015}).
We note, however, that
radio jet emission occurs on scales larger than our computational domain (${>} 10^5 r_\mathrm{g}$; with break frequency $10^{11}-10^{13}$~Hz, \citealt{koljonen_jet_2025}). 
As shown with crosses in Fig.~\ref{fig:Ljet}, accretion columns imply both high jet luminosity, due to the flow penetrating deep inside the light cylinder, and high X-ray luminosity due to the X-ray-emitting hot accretion columns (Appendix~\ref{app:mode_combi}).
Hence, frequent occurrences of accretion columns are inconsistent with the observations, if one assumes that the radio flux is produced by the jet.

\subsection{Accretion torques}
\label{sec:torques}

We compute the total torque acting on the NS, including the gas accretion (due to rotating material crossing the the NS surface) and electromagnetic (due to twisted magnetic field lines) contributions; see Appendix~\ref{app:torques} for more details.
First, we note that the torque is almost exclusively electromagnetic (Fig.~\ref{fig:app_torques}).
We find the strongest torques when the disk is inside the light cylinder.
In the \lowmdot{}\texttt{-w} model (presumably corresponding to the high-$\Lx$ mode), the torque remains at its isolated value, denoted $N_0$, with $N\, {=} \, N_0 \, {\sim} \, \mu^2 \Omega_\star^3 / \mathrm{c^3}$.  
Meanwhile, in the \highmdot{}\texttt{-p} model, we get an averaged torque $N \, {\approx} \, 1.3\, N_0$.
Considering, as previously, $70$\% propeller and $30$\% wind regime during the low mode, this would be $N \, {\approx} \, 1.21\, N_0$.
The increase in the spin-down rate, generally denoted $\dot{P}$, is directly proportional to this torque.
Because J1023 spends  ${\sim}20\text{--}30\%$ of the time in the low mode (e.g., \citealt{linares_x-ray_2014}), the increase in the spin-down rate above the isolated value extrapolated from our analysis would be: (i)~${\sim}6\text{--}9\%$ for a propeller-only low mode and (ii)~${\sim}4\text{--}6\%$ for a $70\%\text{--}30\%$ propeller--wind low mode.
This picture is qualitatively consistent with the results of \cite{burtovoi_spin-down_2020} and \cite{jaodand_timing_2016}, who reported mode-averaged spin-down rates ${\sim}5\%$ and ${\sim}27\%$ higher than in the radio pulsar state, respectively, consistent with a weak spin-down enhancement.

\section{Conclusions}
\label{sec:ccl}

We presented a computational study of the interactions between the pulsar magnetosphere/wind and accretion flow. 
For this, we used a 2D axisymmetric hybrid force-free–GRMHD model. 
We post-processed it with a Monte Carlo radiative transfer code. Our main goal was to investigate the origin of low and high X-ray modes in transitional millisecond pulsars (tMSPs). 
We did this by varying the disk density, and hence the mass inflow rate, that eventually determines the ability of the accretion flow to penetrate the magnetosphere. 
We reproduced two distinct regimes. 
In the first one, associated with low density, the flow is truncated outside the light cylinder by the pulsar wind (Sec.~\ref{sec:res_1}). 
We refer to this as the wind regime. 
In the second one, associated with high density, the flow penetrates the magnetosphere and is mostly located between the corotation radius and the light cylinder radius. 
This regime is similar to the classical propeller regime. It is accompanied by occasional expulsions of the flow to outside the light cylinder.

We estimated and compared the electromagnetic energy dissipation in the wind regime against that of an isolated pulsar (App.~\ref{app:dissipation}).
The wind of an isolated pulsar concentrates in the equatorial plane, where the reconnection in the current sheet then dissipates it \citep{tchekhovskoy_time-dependent_2013}. 
In the wind regime, the conversion of electromagnetic energy into thermal energy is about twice as large near the disk inner edge location, compared to the isolated case. 
This means that the disk is heated as a consequence of wind--disk collision. 
The heated, magnetized (from $\sigma \, {<} \, 1$ up to $\sigma \, {\sim}\,  10$) plasma around the disk surface emits through thermal synchrotron in the X-ray band. 
The X-ray spectrum reaches the high X-ray mode levels and has a comparable spectral slope (Sec.~\ref{sec:NS_mag_field}), for a reasonable choice of the NS magnetic field normalization 
($B_\star\, {\sim} \, 7.9 \times 10^7$~G). 
We found that this X-ray emission is reprocessed, through inverse Compton scattering, up to gamma-ray energies.
This picture is qualitatively consistent with the high X-ray mode of tMSPs (as suggested by \citealt{papitto_pulsating_2019}, \citealt{veledina_pulsar_2019}).
Thus, our simulations reveal a high X-ray mode associated with the wind regime, where the disk is held and heated outside the light cylinder by the pulsar wind.
The origin of the wind regime resides in a low density/inflow rate.

At high density/inflow rates, the gas enters the light cylinder.
Hence, the dissipation and X-ray luminosity drop. 
In this propeller regime, the X-ray flux is smaller than the low X-ray mode values. 
When gas is occasionally expelled from the light cylinder, the system shortly experiences the wind regime though. 
Since our simulations span shorter timescales than typical observations, we investigated whether a combination of these regimes could reproduce the low X-ray mode (App.~\ref{app:mode_combi}).
Indeed, the observed spectral slope \citep{linares_x-ray_2014} is similar in both modes and therefore suggests a common heating mechanism, suggested here to be the wind--induced disk heating, albeit at lower efficiency in the low X-ray mode.
We show that a mix of propeller ($70\%$) and such wind ($30\%$) regimes reconciles the flux level and spectral slope of the model with the data.
The question would then be to explain why this would happen at the fixed rate, found to be $70\%-30\%$ here, needed to reproduce the observed mode stability. 
To circumvent this question, we suggest an alternative model to our axisymmetric model with a varying disk truncation radius. 
3D non-axisymmetric accretion (e.g., \citealt{romanova_threedimensional_2003}, \citealt{2024ApJ...975...57P}) would allow the accretion flow to enter the light cylinder over some fraction of the azimuthal angle, plausibly resulting in wind—induced disk heating with lower efficiency than when the flow is fully outside the light cylinder (i.e. the high X-ray mode configuration).
It could reproduce the low X-ray mode flux and spectral slope. 
Finally, it would lead to an increased NS spin-down rate, in agreement with observational constraints. 
In another alternative scenario, one could associate the low X-ray mode with a disk located even further away from the pulsar; however, this would not produce the observed spin-down rate enhancement \citep{jaodand_timing_2016}.
Hence, our simulated 2D axisymmetric and tentative 3D non-axisymmetric pictures invoke pulsar wind--induced disk heating as the main source of X-ray emission in the low and high X-ray modes.
We also stress the importance of the disk location with respect to the light cylinder.

We report higher jet luminosity in the low X-ray mode configuration and lower jet luminosity in the high X-ray mode configuration (Sec.~\ref{sec:anticor_jet_X}).
This trend is due to the increased open magnetic flux in the propeller regime, assumed to be at work during most of the low X-ray mode epoch.
This result is consistent with the observed X-ray–radio anti-correlation if the radio flux is assumed to originate from the jet. 
Moreover, we find a non-zero but marginal mode-averaged increase of the spin-down rate compared to the isolated/radio pulsar value (driven by increased torques in the low X-ray mode; App.~\ref{app:torques}), in agreement with the observational constraints. 
Hence, our overall scenario --- the low X-ray mode when the flow is mostly inside the light cylinder and the high X-ray mode when it is outside --- is internally consistent.

Optical and X-ray pulsations at the NS spin period are observed in the X-ray high mode \citep{papitto_transitional_2022}.
3D simulations breaking axisymmetry with an oblique pulsar (\citealt{2024ApJ...961L..20M}, \citealt{2024ApJ...960L..12D}) are needed to produce pulsations (e.g., \citealt{veledina_pulsar_2019}).
Therefore, we defer a dedicated study of pulsations, and more generally, of 3D models, to future works.

Our proof-of-concept work underestimates the X-ray cut-off frequency and the gamma-ray flux. 
Future means of resolving these discrepancies include varying the proton-to-electron temperature ratio (e.g., \citealt{moscibrodzka_coupled_2013}) and accounting for non-thermal distributions of electrons. 
Indeed, we considered a thermalized electron-proton plasma throughout, even in high-magnetization regions. 
In reality, the magnetospheric current sheet’s plasmoids and the wind– disk interface are both expected to comprise non-thermal electron-positron pair plasma and contribute to the emission. 
We leave these non-trivial aspects and the exploration of non-thermal emission excited by shocks and reconnection for future work. 
By presenting quantitative physical descriptions for the low and high X-ray modes, which presumably correspond to stable configurations, our study also lays the foundation for studying more transient phenomena causing the sporadic flare mode ($\Lx \, {\sim} 10^{34} \, \mathrm{erg \, s^{-1}}$).
This can last from tens of seconds to hours \citep{bogdanov_coordinated_2015}, which is still longer than the duration of our simulations.

Overall, we presented the first evidence based on GRMHD and radiative transfer calculations that the mass density or equivalently, inflow rate, can control the X-ray modes by varying the pulsar wind--induced disk heating efficiency.
We showed that this model already successfully satisfies, under reasonable assumptions, the observational constraints on the NS magnetic field strength, spin-down enhancement and anti-correlation between X-ray and radio luminosities.
\begin{acknowledgments}
RMR thanks K. Chatterjee, K. Koljonen, M. Mo\'scibrodzka, A. Papitto, V. Richard-Romei and A. Veledina for useful discussions.
We acknowledge funding from the European Research Council (ERC) under the European Union Horizon 2020 research and innovation programme (grant agreement number No. 101002352). 
RMR acknowledges the project Ref. PID2024-157196NB-I00 (MICIU/AEI/10.13039/501100011033).
KP acknowledges support from the Laboratory Directed Research and Development Program at Princeton Plasma Physics Laboratory, a national laboratory operated by Princeton University for the U.S.\ Department of Energy under Prime Contract No.\ DE-AC02-09CH11466. 
AT acknowledges support by NASA
80NSSC26K0343, 
80NSSC22K0031, 
80NSSC22K0799, 
80NSSC18K0565 
and 80NSSC21K1746 
grants, and by the NSF 
AST-2009884, 
AST-2107839, 
AST-1815304, 
AST-1911080, 
AST-2206471, 
AST-2407475 
grants.

\end{acknowledgments}

%






\appendix

\section{Dissipation radial profile}
\label{app:dissipation}

\begin{figure}
\begin{center}\includegraphics[width=\minof{0.75\textwidth}{\columnwidth}]{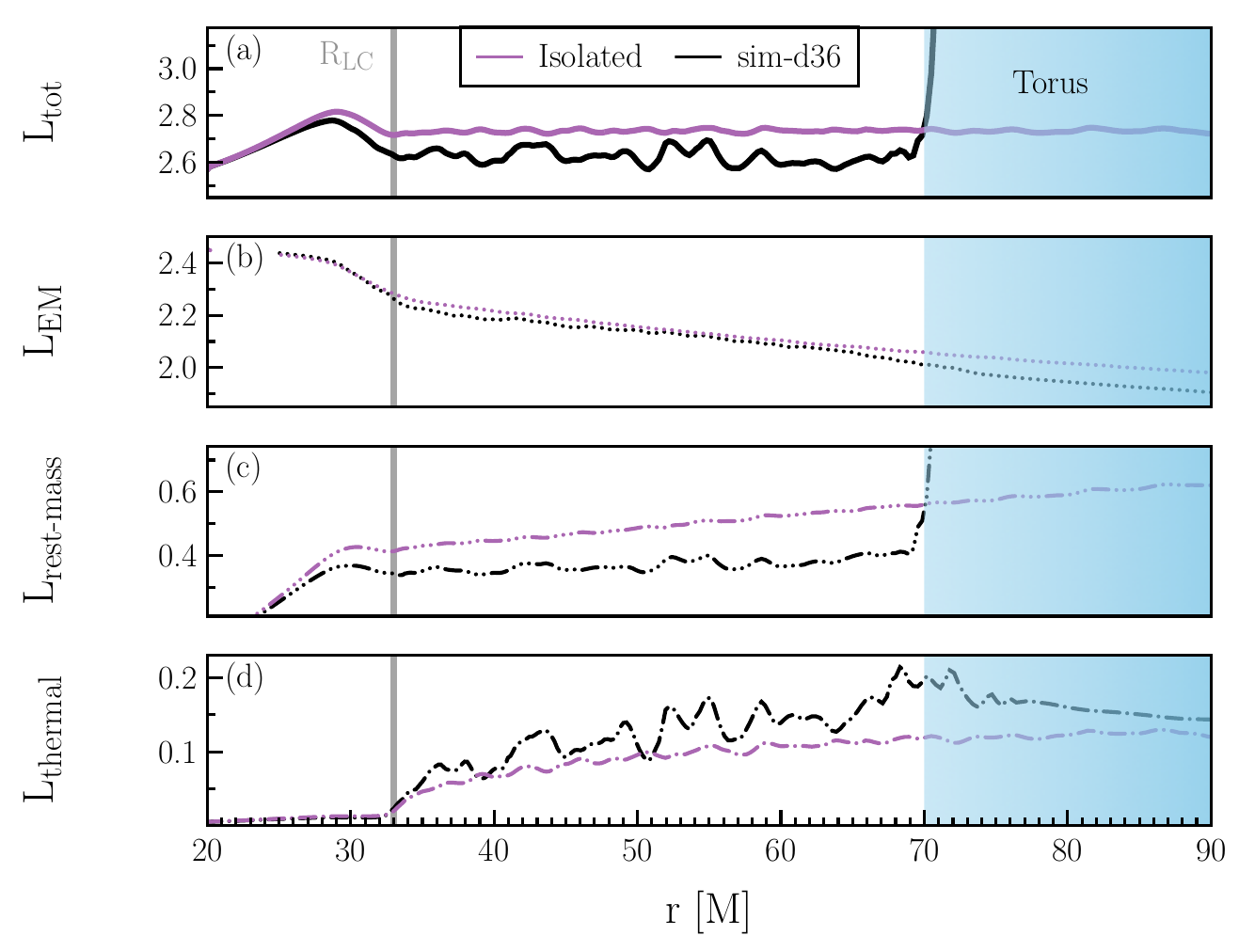}\end{center}
\caption{
Pulsar wind--disk collision leads to enhanced pulsar wind power dissipation, as seen through a larger value for $L_\mathrm{thermal}$ in presence of a torus located outside $\Rlc$ (black curves) than for an isolated pulsar (purple curves) on panel (d).
We show the total luminosity (\textbf{panel a}), the electromagnetic luminosity  (\textbf{panel b}), the rest-mass luminosity (\textbf{panel c}) and the thermal luminosity (\textbf{panel d}).
The black curves correspond to the early phase of the \lowmdot{} simulation (time-averaged over $t \, {\in} \, [1,000;2,000] \, r\mathrm{_g/c}$ to spatially distinguish the torus energy content and the purple curves correspond to the isolated pulsar case.}
\label{fig:dissip_vs_r}
\end{figure}

To understand the heating source (see Fig.~\ref{fig:rho_temp_low_high}c), we compute as integrals over spherical shells the total luminosity 
$L_\mathrm{tot} \, {\equiv} \, - \int T^r_t \sqrt{-g} \mathrm{d} \theta \mathrm{d}\phi$ and the individual contributions as the (wind) electromagnetic luminosity $L_\mathrm{EM} \, {\equiv} \, - \int (b^2 u^r u_t - b^r b_t) \sqrt{-g} \mathrm{d} \theta \mathrm{d}\phi $,
the rest-mass luminosity 
$L_\mathrm{rest-mass} \, {\equiv} \, - \int \rho \mathrm{c^2} u^r u_t \sqrt{-g} \mathrm{d} \theta \mathrm{d}\phi $ and the thermal luminosity $L_\mathrm{thermal} \, {\equiv} \,  - \int (e+p) u^r u_t  \sqrt{-g} \mathrm{d} \theta \mathrm{d}\phi$. 
We recall that the internal energy and pressure are related via our ideal gas equation of state.

Figure~\ref{fig:dissip_vs_r} (panel a) shows the nearly perfect conservation of $L_\mathrm{tot}$ (Eq.~\ref{eq:eqs}) with respect to $r$ beyond $\Rlc$ for the isolated pulsar (purple curve) or between $\Rlc$ and the torus (black curve), which is delimited by the blue region (see Fig.~\ref{fig:rdisk}a).
As shown on panels (b), (c), and (d), $L_\mathrm{EM}$ decreases with radius while $L_\mathrm{rest-mass}$ and $L_\mathrm{thermal}$ increase.
Quantatively, most conversion occurs between $L_\mathrm{EM}$ and $L_\mathrm{rest-mass}$, indicating magnetic acceleration of the plasma.
Nevertheless, the torus deflects the field lines \citep{parfrey_accreting_2024} and thereby reduces the efficiency of the magnetic acceleration and $L_\mathrm{rest-mass}$, compared to the isolated pulsar.
The rest of the electromagnetic luminosity loss is transmitted to the thermal luminosity.
For the isolated pulsar, we attribute it to reconnection in the equatorial current sheet, where plasmoids are visible and suggest plasmoid-mediated reconnection as in the \lowmdot{} simulation. 
This thermal luminosity is constant over ${\sim}\, 100 \, r_\mathrm{g}$ scales and slightly decreases over larger radii, as the plasmoids as less resolved by the hyper-exponential radial spacing.
In presence of the torus, $L_\mathrm{thermal}$ is almost twice as large as in the isolated pulsar case near the torus inner edge radius; we attribute the difference to the wind--disk interaction and to the fraction of the wind luminosity intercepted by the disk.
We get a maximal value of $L_\mathrm{thermal}/L_\mathrm{tot} \, {\approx}\, 6.1 $\%, which, for $L_\mathrm{tot}$ set to realistic isolated pulsar wind powers, or to $L_\mathrm{jet}$ values taken from our simulations (Fig.~\ref{fig:Ljet}), is sufficient to power the X-ray flux observed in the high mode.


\section{Contributions of propeller, wind, and accretor regimes to the modes}
\label{app:mode_combi}

\begin{figure}
\centering
\includegraphics[height=0.34\textwidth,trim=0 0 0.5cm 0,clip]{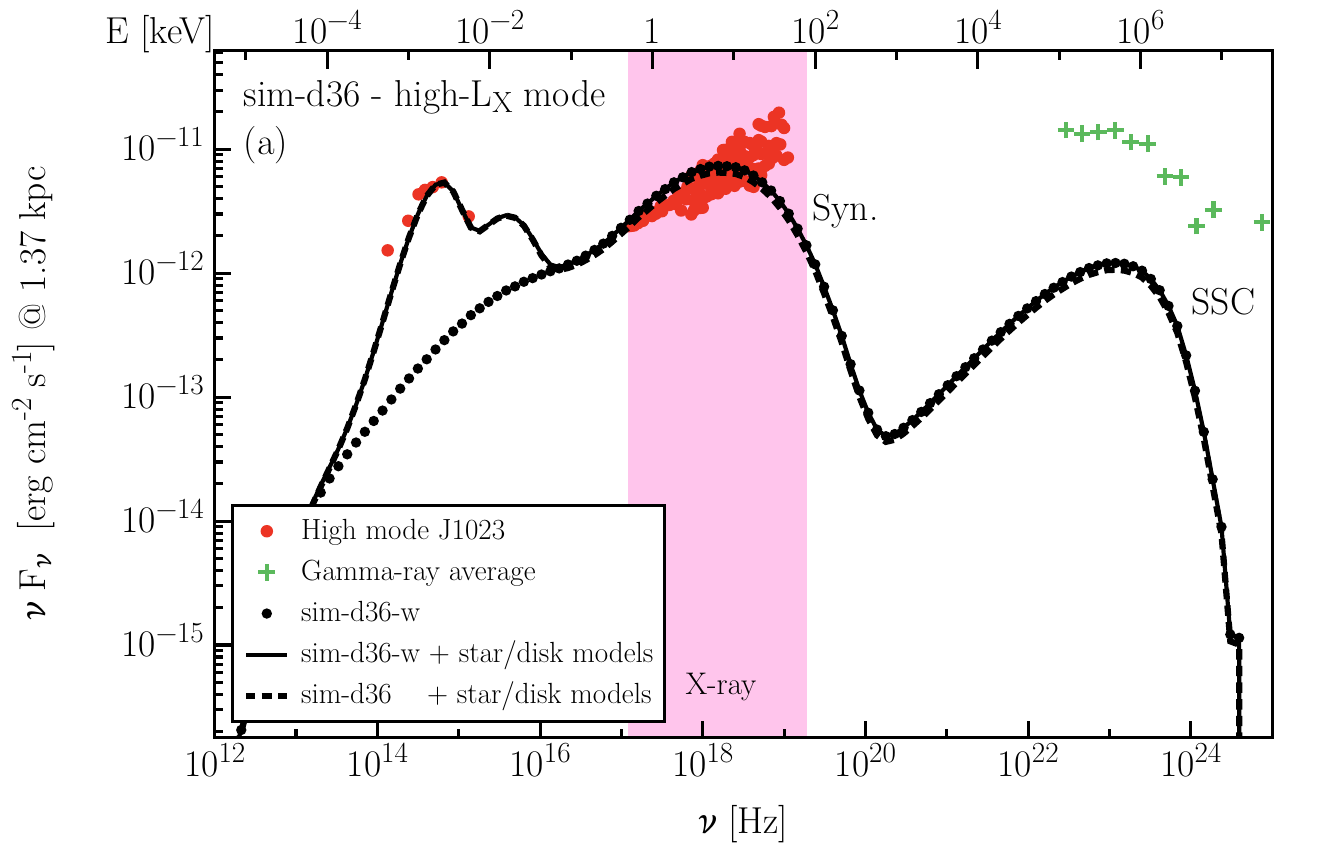}
\includegraphics[height=0.34\textwidth,trim=2.2cm 0 0.5cm 0,clip]
{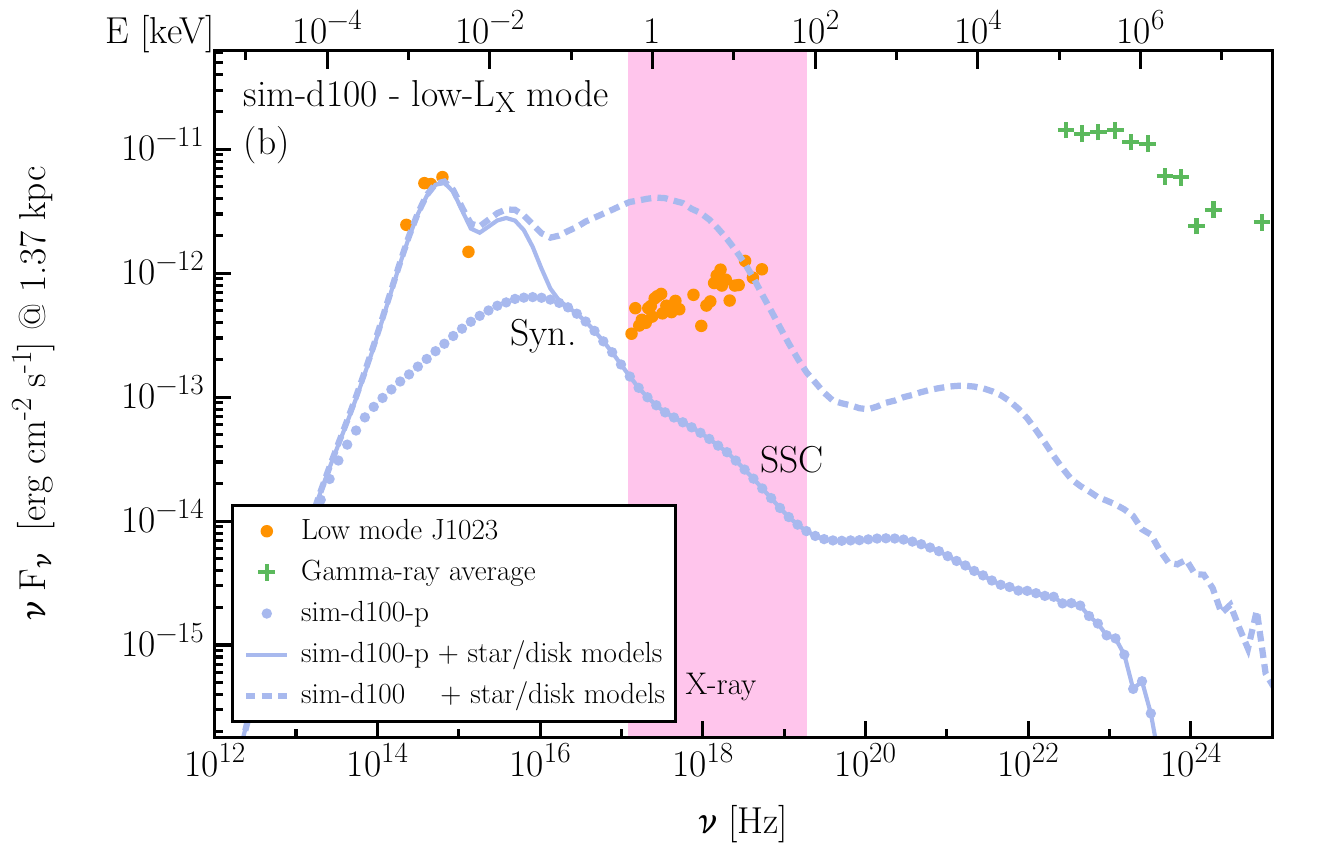}
\caption{
Spectral energy distributions 
from the \lowmdot{} model we compare with the high mode (\textbf{panel a}) and from the \highmdot{} model we compare with the low mode (\textbf{panel b}).
Contributions include the time-averaged GRMHD dominant regime (propeller or wind) with the additional contributions from the optical companion and thermal, thin disk model (full line) and without them (dots).
The SED averaged over the entire post-processed epoch is shown for comparison (dashed line).
We indicate the dominant emission mechanism contributing to different parts of the SEDs; on panel (b), this corresponds to the propeller regime.
X-ray (orange/red dots) and gamma-ray (green pluses; not tied to a mode) data, taken from \citealt{baglio_matter_2023} and \citealt{papitto_pulsating_2019}, respectively, are plotted for comparison.
The large X-ray band (0.5-80~keV) is shown in pink.}
\label{fig:app_SED_low_high}
\end{figure}

In this Appendix, we describe in detail the individual contributions leading to the SEDs shown in Fig.~\ref{fig:SED_low_high}.
We show in Fig.~\ref{fig:app_SED_low_high} the more representative GRMHD model (\highmdot{}\texttt{-p} and \lowmdot{}\texttt{-w}, respectively) with 
the star and disk models (solid lines) and without them (dots).
These star and disk components are included to account for the optical and UV data.
The dashed lines refer to SEDs averaged over the entire post-processed epochs, given in Sec.~\ref{sec:res_1} (and Fig.~\ref{fig:rdisk}), including those times with the flow either penetrating the light cylinder (in the \lowmdot{} case), accreting through columns or being expelled from the light cylinder (in the \highmdot{} case).
These longer timescales help us to study how some departure from idealized configurations impact the SED.
For example, we want to estimate the impact of the few and short occurrences of propeller regime in \lowmdot{}, which is in the wind regime most of the time.
These changes still occur on much shorter timescales than typical exposure times.
For the \lowmdot{} model, the dashed and solid lines are hardly distinguishable because the time-averaged flux is largely dominated by the contribution produced in the \lowmdot{}\texttt{-w} regime.
For the \highmdot{} model, the flux is higher in the wind regime and in the accretor regime compared to the propeller regime.
The resulting time-averaged, synchrotron-powered, X-ray emission overshoots the observed soft X-ray, low mode flux.

\begin{figure}
\includegraphics[width=\columnwidth]{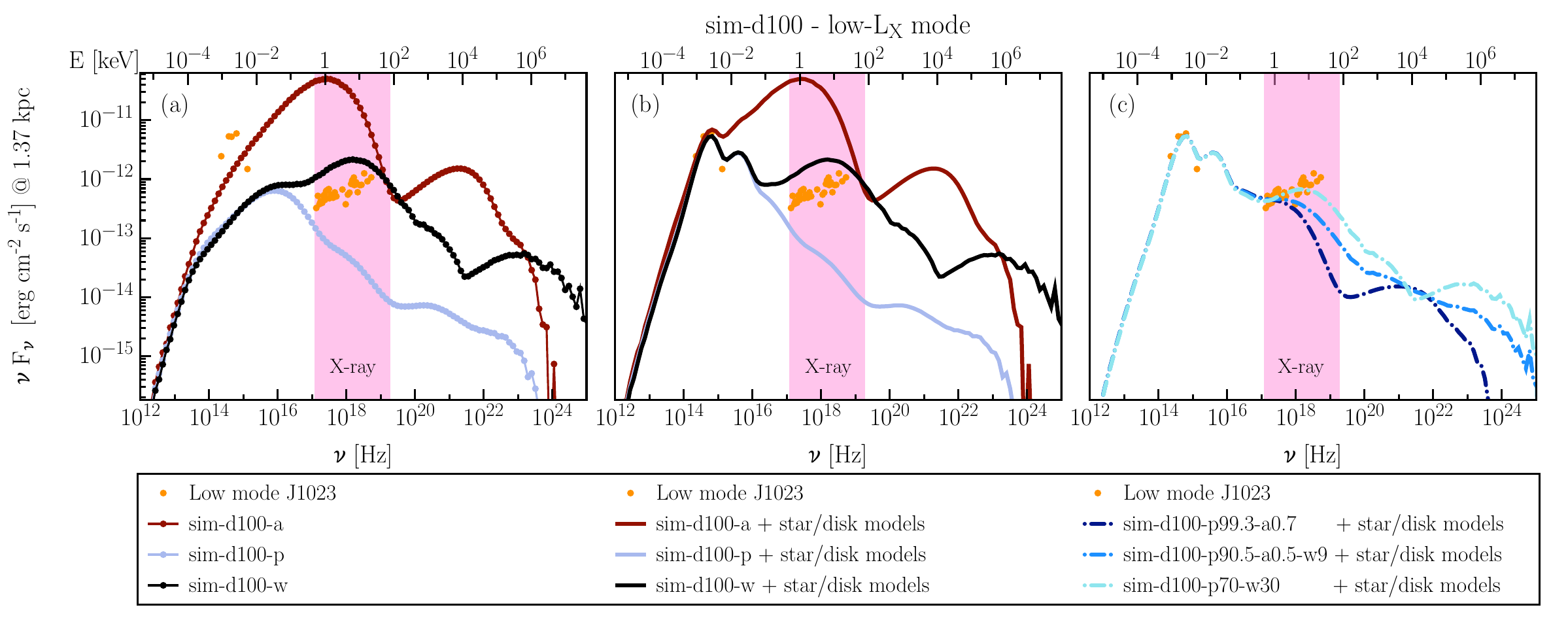}
\caption{
Combinations of propeller and wind regimes in the \highmdot{} model could reproduce the observed low mode data.
\textbf{(panel (a))}: low mode J1023 data (orange dots) and individual SEDs of the accretor, propeller, and wind ($\Rm \, {>} \, \Rlc$) regimes.
\textbf{(panel (b))}: low mode J1023 data (orange dots) and full SEDs (GRMHD data and star/disk models).
\textbf{(panel (c))}: low mode J1023 data (orange dots) and combinations of the full SEDs between the accretor, propeller, and wind regimes.
The data are best reproduced with the combination of $70\%$ propeller and $30\%$ wind regimes (\highmdot{}\texttt{-p70-w30}).}
\label{fig:app_SED_low_mode_combi}
\end{figure}

Summing up, \lowmdot{}\texttt{-p} underestimates the low mode X-ray flux, and \lowmdot{} overestimates it. Because of this, we investigate below if weighted combinations of the three dynamical regimes---propeller, accretor, and wind---could reproduce the low mode data.
First, we show the individual averaged SEDs over each of these three regimes in panel (a) of Fig.~\ref{fig:app_SED_low_mode_combi}.
It can be seen that the system is much brighter in the accretor regime than the others.
The SED peaks at lower energy (a few $10^{17}$~Hz) in the accretor regime than the wind regime (${\sim}10^{18}$~Hz).
This wind regime produces a similar SED slope to the simulated high mode in the \lowmdot{} model (Fig.~\ref{fig:app_SED_low_high}, right panel), with slightly lower flux ($2 \times 10^{12} \, \mathrm{erg \, cm^{-2} \, s^{-1}}$ against $6 \times 10^{12} \, \mathrm{erg \, cm^{-2} \, s^{-1}}$ at $10^{18}$~Hz).
In Fig.~\ref{fig:app_SED_low_mode_combi}(b), we show the GRMHD data with the additional star and disk models.
It can be seen that the accretor regime largely overshoots the UV and X-ray data.
Meanwhile, the wind regime produces a flux slightly above the X-ray data, by a factor of ${\sim}\,3$, with a similar slope. This suggests a contribution --- quantified below --- from pulsar wind--induced disk heating in the low mode.

We study here how various weighted combinations of these regimes, akin to time averages, could help to reproduce the low mode data.
Because this model exhibits mostly a propeller regime, we will start by assuming that the propeller epoch is the longest.
Since the accretor regime overshoots the X-ray flux at ${\sim} \, 1$~keV in particular, we first adjust these models to fit at this energy.
The results are shown in Fig.~\ref{fig:app_SED_low_mode_combi}(c).
The blue curve corresponds to a combination of propeller ($99.3$\%) and accretor regimes ($0.7$\%).
It shows that the accretor regime cannot represent more than $0.7$\% during the low mode. 
The SED slope is also incompatible with the observed data.
Instead, introducing a wind regime contribution ($9$\%, dark green curve) improves the slope and requires us to reduce the accretor contribution to $0.5$\%, to not overestimate the soft X-ray flux.
Finally, the low mode SED data are best reproduced with $30$\% wind regime and $70$\% propeller (light green curve), i.e., without accretion columns; this is our best model for the low mode (Fig.~\ref{fig:SED_low_high}). This fraction of wind regime leads to a spectral slope compatible with the data.

\section{Torques}
\label{app:torques}

\begin{figure}
\begin{center}
  \includegraphics[width=\minof{0.75\textwidth}{\columnwidth}]{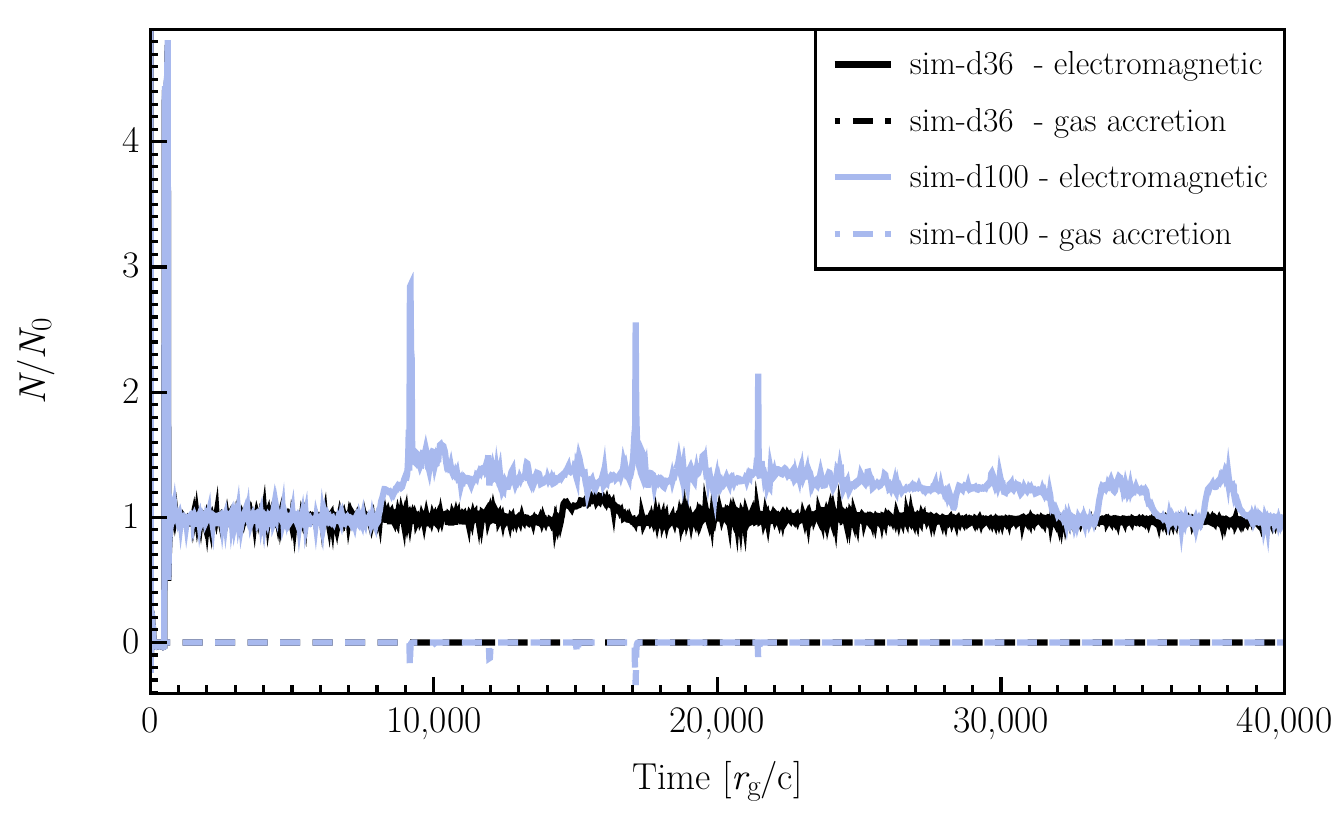}
\end{center}
\caption{Time evolution of the gas accretion (dashed line) and electromagnetic (full line) torque applied to the NS, normalized to the isolated value, in models \lowmdot{} (black) and \highmdot{} (light blue).}
\label{fig:app_torques}
\end{figure}

We show in Fig.~\ref{fig:app_torques} the (gas accretion and electromagnetic) torques $N$, normalized to the isolated value $N_0$, exerted on the NS as a function of time in both models.
First, it can be seen that the gas accretion torque is generally negligible compared to the electromagnetic one.
During rare gas accretion episodes, the accretion and electromagnetic torques reach a maximal value of $0.2 \, N_0$ and ${\approx} \, 3\, N_0$ in amplitude, respectively.
After an initial phase ($t \, {<} \, 500\, r_\mathrm{g}/\mathrm{c}$) coincident with zero NS rotation and zero torques, the electromagnetic, or total, torque increases dramatically as the magnetosphere relaxes, then settles to its isolated value.
Later in time, the increases in the torque value coincide with the accretion flow penetrating the light cylinder and settling into a propeller regime.
In the \highmdot{} model (low mode), we find a time-averaged torque in the propeller regime of $N \, {\approx} \, 1.3\, N_0$ (see Sec.~\ref{sec:torques} for discussion).


\bibliography{Zotero.bib}{}
\bibliographystyle{aasjournal}



\end{document}